\documentclass[aps,rmp,twocolumn,superscriptaddress,nofootinbibt,longbibliography]{revtex4-2}
\pdfoutput=1
\usepackage[utf8]{inputenc}
\usepackage[english]{babel}
\usepackage[T1]{fontenc}
\usepackage{amsmath}
\usepackage{hyperref}

\usepackage{tikz}
\usepackage{lipsum}

\usepackage{amssymb}
\usepackage{amsfonts}
\usepackage{epsfig,pstricks,graphicx}
\usepackage{bm}
\usepackage{physics}
\usepackage{amsthm}
\usepackage[shortlabels]{enumitem}
\usepackage{ulem}

\usepackage{color}
\usepackage{array}
\def\id{{\rm 1\kern-.22em l}}

\begin{document}

\title{Relativistically invariant encoding of quantum information revisited}

\author{Konrad Schlichtholz}
\affiliation{International Centre for Theory of Quantum Technologies (ICTQT),
University of Gdansk, 80-309 Gdansk, Poland}
\author{Marcin Markiewicz}
\affiliation{International Centre for Theory of Quantum Technologies (ICTQT),
University of Gdansk, 80-309 Gdansk, Poland}

\begin{abstract}
In this work, we provide a detailed analysis of the issue of encoding of quantum information which is invariant with respect to arbitrary Lorentz transformations. We significantly extend already known results and provide compliments where necessary. 
In particular, we introduce novel schemes for invariant encoding which utilize so-called pair-wise helicity -- a physical parameter characterizing pairs of electric-magnetic charges. We also introduce new schemes for ordinary massive and massless particles based on states with fixed total momentum, in contrast to all protocols already proposed, which assumed equal momenta of all the particles involved in the encoding scheme. 
Moreover, we provide a systematic discussion of already existing protocols and show directly that they are invariant with respect to Lorentz transformations drawn according to  any distribution, a fact which was not manifestly shown in previous works.
\end{abstract}

\maketitle

\section{Introduction and Motivation}

Almost all quantum information processing protocols can be divided into three stages: (i) state preparation, which encodes some initial information in a multipartite quantum system, (ii) state transformation, which can include controllable (implementing some algorithm or procedure) and uncontrollable (external noise of different origin) operations, and (iii) final measurement. In many cases, it is desirable that the initial encoding of quantum information should be invariant with respect to some class of uncontrollable state transformations.
This is possible if and only if the class of transformations  acts  on the state space of a multipartite physical system such that irreducible representations of this action occur with non-zero multiplicities, which means that \textit{equivalent} irreducible representations act on \textit{different} subspaces of the state space.
Such a condition is fulfilled for example if  the class of transformations includes collective operations, which act in the same way on each of the three or more particles. The arising invariant states belong to either so called Decoherence-Free Subspaces (DFSp) or Decoherence-Free Subsystems (DFSys) \cite{Zanardi97, Bartlett03a}, which we are going to define precisely in the next section. A representative example of invariant encoding of quantum information is the one based on permutation operators acting on many-qubit states, which are invariant under the collective action of single-qubit unitary operators \cite{Bartlett07}. 

The problem of invariant encoding is especially interesting in the context of scenarios in which it is challenging to establish a common reference frame (RF) shared between the party involved in stage (i), which performs encoding of information into the initial state, and parties performing actions on the state: stages (ii) and (iii).  Then one can consider troublesome transformations of RFs that make establishing a common RF challenging and prepare the state using DFSp or DFSys that would be protected against these transformations. Such a scenario could be encountered, e.g., when information is transmitted over long distances in space or when one stores quantum information in memory and performs its readouts on spaceship which undergoes boosts. In particular, one can expect relativistic effects to affect the quantum information protocols \cite{Peres04}. The primary example of the reasons for this is the fact that the density matrix of spin-$\frac{1}{2}$  particle is not covariant under Lorentz transformations \cite{Peres03,Spin_cov2}.  This gave rise to the issue of relativistically invariant encoding which was first addressed in \cite{Bartlett05} for the special case of uniform distribution of Wigner's rotation angles related with Lorentz transformations on spin degrees of freedom and for equal sharply defined momentum of all particles used in encoding information. Since then, relativistic quantum information has been broadly considered \cite{RQI_1,RQI_2,RQI_3,RQI_4,RQI_5,RQI_6,RQI_7}. However, in the article \cite{Dragan15} authors noted that the research concentrated on completely uncorrelated RFs, and analyzed the efficiency of communication with partially correlated RFs. Here, uncorrelated RFs stand for the scenarios in which one uses a uniform distribution of group transformations.  This is in fact a good avenue for this subfield as considerations of fully uncorrelated RFs resulting from Lorentz transformations are operationally meaningless, as one always has the knowledge that his ''space ship`` is not boosting during execution of the protocol in random directions with velocities arbitrarily close to the speed of light. Furthermore, invariant encoding in terms of DFSp and DFSys can also be utilized in partially correlated RFs scenarios. However, the fundamental papers \cite{Bartlett05,Bartlett07} for relativistic quantum information either skip the proof or utilize Schur's Lemma, which can be used only if the transformations are drawn according to the group invariant measure  i. e. it only applies to the  completely uncorrelated RFs. What is more, in the case of non-compact groups the group invariant measure is  globally infinite (it is finite only on compact subsets of transformations), which further complicates the analysis of  completely uncorrelated RFs in the relativistic context.

Another potentially relevant development for relativistically  invariant encodings is  the fact that recently a new full classification of  multiparticle representations of Poincar\'e group was developed  \cite{Csaki21}. This classification includes previously omitted degree of freedom which appears in multi-particle states, that is pair-wise helicity.  While pair-wise helicity appears trivially in most circumstances, its transformation properties have an impact on states of systems involving electric and magnetic charges. Although magnetic monopoles have not yet been observed as elementary particles, they appear as topological quasiparticles in condensed matter systems \cite{Monopoles_exp,Monopoles_cond}. Therefore, for completeness one should factor in pair-wise helicity into the considerations of relativistically  invariant encoding of quantum information. This also opens new possibilities, as additional degrees of freedom appearing in the system could allow for denser encoding of the information with the same number of particles.

In this paper, we extend the  theory of relativistically invariant encoding with equal momenta to include all affected by Poincar\'e group degrees of freedom. This accounts for pair-wise helicity which contributes additional non-trivial degrees of freedom in systems containing electric-magnetic charges. Upon this, we propose a simple encoding scheme using pair-wise helicity that allows for denser encoding. Furthermore, we propose relativistically invariant encoding which does not require the momenta of all particles in the system to be equal. We also show that encoding with equal momentum particles is a special case of our new encoding. In addition, we provide a pedagogical and rigorous proof of the existence of DFSp and DFSys in the general scenario of any distribution of transformations applied to the state space both for compact and non-compact groups as long as these transformations act via unitary representations.

This work is structured as follows: Section \ref{sectionI} provides the general theory of invariant encoding of quantum information. Section \ref{sectionII} reviews behaviour of multi-particle states under Poincar\'e transformations. Section \ref{sectionIII_col} describes the general theory of equal momentum relativistically invariant encodings, including pair-wise helicity. Section \ref{sectionIII_non_col} introduces non-equal momentum relativistically invariant encoding. Finally, Section \ref{conclusions} presents the conclusions of this article.





\section{Invariant encoding of quantum information in a nutshell}\label{sectionI}

\subsection{General case}

Let us assume that a multipartite quantum system initially prepared in a state $\rho$ is subject to a class of uncontrollable transformations acting on the state space via some representation $\pi_{\mathcal G}$ of a symmetry group $\mathcal G$. We define a general \textit{twirling} transformation with respect to a group $\mathcal G$ by:
\begin{equation}
    \label{def-twirl-gen}
    \mathcal T_{\mathcal G}(\rho)=\int_{\mathcal G}\operatorname{d}g\,f(g)\pi_{\mathcal G}(g)\rho\,\pi_{\mathcal G}(g)^{\dagger},
\end{equation}
in which the function $f$ is a weight function (in particular we allow $f$ to be a Dirac delta $f(g)=\delta(g-g_0)$ for some group element $g_0$). The integral $\int_{\mathcal G}$ is performed over the entire group manifold and we assume  normalization condition $\int_{\mathcal G}\operatorname{d}g\,f(g)=1$.
For the sake of further clarity let us explain the notation $\pi_{\mathcal G}(g)$. $\pi_{\mathcal G}$ denotes a representation, that is, a mapping from the group $\mathcal G$ to linear operators on some representation space. Here, the representation space is some fixed separable Hilbert space $\mathcal H$, possibly of infinite dimension. $\pi_{\mathcal G}(g)$ is the image of this mapping for the fixed group element $g$, which is some concrete linear operator. Finally, $\pi_{\mathcal G}(\mathcal G)$ is the entire operator algebra generated by operators $\pi_{\mathcal G}(g)$.
We demand that the map \eqref{def-twirl-gen} is positivity-preserving and trace-non-increasing, as otherwise it cannot be interpreted as a proper quantum transformation.

We say that an encoding of quantum information is invariant with respect to the action on the state space of a symmetry group $\mathcal G$, if it is realised via states $\rho_{\mathcal G}$ invariant with respect to the map \eqref{def-twirl-gen}:
\begin{equation}
    \label{def-inv-state}
    \rho_{\mathcal G}=\mathcal T_{\mathcal G}(\rho_{\mathcal G}),
\end{equation}
for \textit{any choice of the measure $\operatorname{d}g$ and the weight function $f(g)$}.
A class of such states is non-trivial if and only if $\pi_{\mathcal G}$ decomposes into irreducible representations with non-trivial multiplicities, which means that at least two equivalent irreducible  representations act on different subspaces of the state space.  We will describe this phenomenon more thoroughly in Section \ref{sec:genInvState}, however, firstly we discuss a familiar example.

\subsection{$N$ independent finite-dimensional quantum systems}

Let us take $N$ $d$-dimensional fully distinguishable quantum systems ($N$ qu-$d$-its), whose state space is represented by the tensor product space $\mathcal H=(\mathbb C^d)^{\otimes N}$.
In this case, the problem of invariant encoding is fully solved by means of Schur-Weyl duality \cite{Markiewicz22}. We have to treat separately two cases: 
\begin{itemize}
    \item group $\mathcal G$ is compact, and therefore has unitary representations on $(\mathbb C^d)^{\otimes N}$.
    \item group $\mathcal G$ is non-compact, and therefore all representations of $\mathcal G$ on $(\mathbb C^d)^{\otimes N}$ are non-unitary.
\end{itemize}
Note that the above alternative is valid only in the case of finite-dimensional representations (as is the case of representations on  $(\mathbb C^d)^{\otimes N}$), since non-compact groups have infinite-dimensional unitary representations (see Sec. \ref{sectionII}).
In the first case the invariant encoding always exists in the strict sense \eqref{def-inv-state} as long as at least one  of the irreducible  representations $\{\pi^i_{\mathcal G}\}$ of $\mathcal G$ on $(\mathbb C^d)^{\otimes N}$ appears with multiplicity greater than one: in such a case the 
commutant of $\pi_{\mathcal G}(\mathcal G)$, treated as matrix algebra, with respect to the general linear group $\textrm{GL}(dN, \mathbb C)$ is non-trivial and can be used to encode invariant quantum information.
 In order to illustrate this formal idea let us consider two groups: the unitary group $\textrm{U}(d)$ and the permutation group of $N$ elements $\textrm{S}_N$.
Let us fix two representations of these groups on $(\mathbb C^d)^{\otimes N}$:
\begin{eqnarray}
\label{collU}
&&\pi_{\operatorname{U}(d)}(U)=U^{\otimes N},\nonumber\\
&&\pi_{\operatorname{S}_N}(p)=O_p,
\end{eqnarray}
in which $U$ is just an element of a fundamental representation of $\textrm{U}(d)$ on  $\mathbb C^d$, whereas $O_p$ is an image of a representation of a permutation group of $N$ elements by orthogonal matrices on $(\mathbb C^d)^{\otimes N}$ known as \textit{tensor permutation operators} (see \cite{Christian06, Christian07, Markiewicz22}). Note that $\pi_{\operatorname{U}(d)}$ acts collectively (at the same manner) on all $N$ subsystems, whereas $\pi_{\operatorname{S}_N}$ acts globally (permutes subsystems); nevertheless, both representations are reducible. What is more, their images  $\pi_{\operatorname{U}(d)}(\operatorname{U}(d))$ and 
$\pi_{\operatorname{S}_N}(\operatorname{S}_N)$, treated as matrix algebras, are mutual commutants with respect to the matrix algebra of endomorphisms on $(\mathbb C^d)^{\otimes N}$.  Then states which are in the span of the operator basis of permutation operators ensure encoding of quantum information, which is invariant with respect to the collective unitary noise, whereas states, which are in the span of the operator basis of collective unitary operators allow for encoding invariant with respect to permutation of subsystems. This property in the context of uniform sampling of group elements in \eqref{def-twirl-gen} has been called \textit{duality of averaging} in \cite{Markiewicz22}.

In the second case, when $\mathcal G$ is non-compact, invariant encoding exists only up to postselection onto subspaces invariant with respect to the action of $\pi_{\mathcal K}(\mathcal K)$, in which $\mathcal K$ is the maximal compact subgroup of $\mathcal G$ and $\pi_{\mathcal K}$ is a representation of $\mathcal K$ taken by restricting  $\pi_{\mathcal G}$. Since this aspect is out of the main scope of the paper, we move a more thorough discussion on this topic to the Appendix \ref{app:SLOCC}.

\subsection{General construction of invariant encoding based on multiplicity spaces}
\label{sec:genInvState}
In this section, we provide a general construction of invariant states based on multiplicity spaces of irreducible unitary representations (not necessarily finite dimensional) of some group $\mathcal G$. Although this construction is classical and can be found in several places \cite{Zanardi97, Bartlett07}, we want to present its most general (hence simplest) version devoid of any unnecessary assumptions.
Let us start by introducing a specific basis, often called the Schur basis, elements of which we denote  by $\ket{i,r,\mu}$, in which $i$ numbers an irreducible representation $\pi_{\mathcal G}^i$ of  some symmetry group of transformations $\mathcal G$. The subsets $\{\ket{i,r,\mu}\}_r$ for fixed value of $\mu$ span subspaces irreducible with respect to the action of $\mathcal G$ and $\mu$ plays the role of the multiplicity index, namely subspaces $\{\ket{i,r,\mu}\}_r$ and $\{\ket{i,r,\eta}\}_r$ correspond to equivalent but different irreducible representations. 
The index $r$ just labels consecutive vectors within each irreducible subspace.
 Then the states $\ket{i,r,\mu}$, for a fixed value of $i$, can be organized into the following array:
 \begin{eqnarray}
    \label{SchurBasis}
     L^i_1:\,\,&&\ket{i,1,1}\,\,\,\,\,\,\,\,\,\ket{i,2,1}\,\,\,\,\,\,\ldots\,\,\,\,\,\,\ket{i,D^i_L,1} \nonumber\\
     L^i_2:\,\,&&\ket{i,1,2}\,\,\,\,\,\,\,\,\,\ket{i,2,2}\,\,\,\,\,\,\ldots\,\,\,\,\,\,\ket{i,D^i_L,2} \nonumber\\
     && \ldots\ldots\ldots\ldots\ldots\ldots\ldots\ldots\ldots\ldots\ldots\ldots \nonumber\\
L^i_{D^i_V}:\,\,&&\underbrace{\ket{i,1,D^i_V}}_{V^i_1}\,\,\underbrace{\ket{i,2,D^i_V}}_{V^i_2}\,\,\,\ldots\,\,\,\underbrace{\ket{i,D^i_L,D^i_V}}_{V^i_{D^i_L}}\nonumber\\
    \end{eqnarray}
where $L^i_{\mu}$ denotes subspace spanned by states of $\mu$-th row and $V^i_r$ denotes analogously subspace for $r$-th column.  The entire representation space is a direct sum of the subspaces 
\begin{equation}
    \label{Pi}
    P_i = \bigoplus_\mu L^i_\mu = \bigoplus_r V^i_r
\end{equation}
with respect to the index $i$. The subspaces $L^i_1,\ldots,L^i_{D^i_V}$ correspond to all equivalent irreducible representations of $\mathcal G$.
The number $D^i_L$ represents the dimension of an irreducible representation, whereas $D^i_V$ denotes its multiplicity.
The range of the index $i$ is not specified and in fact it can be countably infinite.

In order to simplify further argumentation, let us introduce the notion of \textit{virtual tensor products} \cite{ZanardiVirt01}. Namely, the vectors $\ket{i,r,\mu}$ can be seen as elementary tensors of the form $\ket{r}_i\otimes \ket{\mu}_i$,  in which vectors $\{\ket{r}_i\}$ and $\{\ket{\mu}_i\}$ span \textit{virtual subspaces} of dimensions respectively $D^i_L$ and $D^i_V$. Let us introduce three other objects, namely let $\hat e_i^{r_1r_2}$ be elements of a standard matrix basis on the space of 
$D^i_L\times D^i_L$ matrices, analogously let  $\hat e_i^{\mu_1\mu_2}$ denote elements of such a basis on the space of $D^i_V\times D^i_V$ matrices, and finally let $\mathcal I_i$ be a trivial immersion of the matrices defined on the $i$-th  subspace $P_i$ \eqref{Pi}  into the space of matrices on the entire representation space spanned by all vectors $\ket{i,r,\mu}$ (which can be countably infinite as an infinite direct sum). Now the operators acting on virtual subsystems $\{\ket{r}_i\}$ and $\{\ket{\mu}_i\}$ are spanned by basis operators of the form $\hat e_i^{r_1r_2}\otimes\id_{D^i_V}$  and respectively $\id_{D^i_L}\otimes\hat e_i^{\mu_1\mu_2}$. It is very convenient to express these two operator bases directly using outer products of the Schur basis vectors $\ket{i,r,\mu}$. For this aim, we define, following \cite{Markiewicz22}, the following outer-product-based operators:
    \begin{equation}
    \label{FullPiBasis}
    \hat\Pi_{ij}^{r_1\mu_1 r_2\mu_2}=\ket{i,r_1,\mu_1}\bra{j,r_2,\mu_2}.
\end{equation}
Note that they are defined on the entire representation space $\bigoplus_i P_i$.
Using the above definition, we define two families of operators:
\begin{eqnarray}
\label{PiBasis}
\hat\Pi^{\mu_1\mu_2}_i&=&\sum_{r=1}^{D^i_L}\hat\Pi_{ii}^{r\mu_1 r\mu_2},\nonumber\\
\hat\Pi^{r_1 r_2}_i&=&\sum_{\mu=1}^{D^i_V}\hat\Pi_{ii}^{r_1\mu r_2\mu}.
\end{eqnarray}
On condition that the basis $\ket{i,r,\mu}$ is ordered column-wise within each $i$-th subspace \eqref{SchurBasis}, they represent operator bases for virtual subsystems immersed into the entire representation space, when seen in the basis $\ket{i,r,\mu}$ ordered as stated (see \cite{Markiewicz22}, Appendix A1):
\begin{eqnarray}
\label{PiMTensor}
&&\hat\Pi_i^{r_1r_2}=\mathcal I_i\left(\hat e_i^{r_1r_2}\otimes\id_{D^i_V}\right)\nonumber\\
&&\hat\Pi_i^{\mu_1\mu_2}=\mathcal I_i\left(\id_{D^i_L}\otimes\hat e_i^{\mu_1\mu_2} \right).
\end{eqnarray}
From the above representation, it is clear that the above operators always commute, as for different irreps index $i$ they act on different blocks of the direct sum, whereas for the same irreps index  they act on different factors of virtual tensor  product:
\begin{equation}
    \label{PiCommute}
    \left[\hat\Pi_i^{r_1r_2}, \hat\Pi_j^{\mu_1\mu_2}\right]=0.
\end{equation}
Now, since the irreducible representation $\pi_{\mathcal G}^i$ acts irreducibly on virtual subsystems  spanned by
$\{\ket{r}_i\}$ and trivially on virtual multiplicity subsystems spanned by  $\{\ket{\mu}_i\}$, its matrix representation is spanned by operators $\mathcal I_i\left(\hat e_i^{r_1r_2}\otimes\id_{D^i_V}\right)$. Therefore, due to the first formula of \eqref{PiMTensor} its matrix representation in basis  $\ket{i,r,\mu}$ ordered as described above has the following simple form:
\begin{eqnarray}
    \label{Ut1}
    \hat\pi_{\mathcal G}^i&=&\frac{1}{D^i_V}\sum_{r_1,r_2=1}^{D^i_L}\operatorname{Tr}\left(\hat\pi_{\mathcal G}^i\hat\Pi^{r_1 r_2\dagger}_i\right)\hat\Pi^{r_1 r_2}_i\nonumber\\
    &=&\frac{1}{D^i_V}\sum_{r_1,r_2=1}^{D^i_L}[\hat\pi_{\mathcal G}^i]_{r_1r_2}\hat\Pi^{r_1 r_2}_i.
\end{eqnarray}
Note that the above is not a definition of $\hat\pi_{\mathcal G}^i$, which appears on both sides, but a statement that $ \hat\pi_{\mathcal G}^i$ can be decomposed solely using operators $\hat\Pi_i^{r_1r_2}$.  Now, we have all the ingredients to construct an invariant encoding.  An invariant state is spanned solely by the $\hat\Pi_j^{\mu_1\mu_2}$ operators, namely:
\begin{equation}
    \label{rhoInvariant}
    \rho_{\mathcal G}=\sum_{i}\frac{1}{D^i_L}\sum_{\mu_1\mu_2=1}^{D^i_V}\rho^{i}_{\mu_1\mu_2}\hat\Pi_{i}^{\mu_1\mu_2}.
\end{equation}
Note that this state is block-diagonal with respect to index $i$ numbering irreducible representations; therefore, it does not contain any coherences between subspaces corresponding to different irreducible representations (existence of such coherences is forbidden by superselection rules \cite{Bartlett07}).
Since $\left(\hat\Pi_{i}^{\mu_1\mu_2}\right)^{\dagger}=\hat\Pi_{i}^{\mu_2\mu_1}$, which follows from \eqref{FullPiBasis} and \eqref{PiBasis}, we demand that in the above expansion the coefficients $\rho^{i}_{\mu_1\mu_2}$ fulfill a relation $\rho^{i}_{\mu_1\mu_2}=\rho^{i*}_{\mu_2\mu_1}$, in which $^*$ denotes a complex conjugate, in order to guarantee the hermiticity of the density matrix.

Now we will prove that the state $\rho_{\mathcal G}$ \eqref{rhoInvariant} is indeed invariant with respect to the action of group $\mathcal G$. 
Let us assume that the representation $\hat\pi_{\mathcal G}=\sum_i\hat\pi_{\mathcal G}^i$ is unitary. 
Note that we use ordinary sum $\Sigma_i$ instead of direct sum operation $\bigoplus_i$ since in our convention representation $ \hat\pi_{\mathcal G}^i$ \eqref{Ut1} is already defined as operator on the entire representation space, despite the fact that it denotes particular $i$-th irreducible representation. 
Consider  a general twirling operation with respect to this representation:
\begin{eqnarray}
    \label{def-twirl-gen-proof}
    \mathcal T_{\mathcal G}(\rho_{\mathcal G})&=&\int_{\mathcal G}\operatorname{d}g\,f(g)\hat\pi_{\mathcal G}(g)\rho_{\mathcal G}\,\hat\pi_{\mathcal G}(g)^{\dagger}\nonumber\\
    &=&\int_{\mathcal G}\operatorname{d}g\,f(g)\sum_i\hat\pi_{\mathcal G}^i\rho_{\mathcal G}\,\sum_j\hat\pi_{\mathcal G}^{j\dagger}.
\end{eqnarray}
Now note that $\rho_{\mathcal G}$ commutes with each $\hat\pi_{\mathcal G}^i$:
\begin{equation}
    \label{RoCommute}
    \left[\rho_{\mathcal G}, \hat\pi_{\mathcal G}^i\right]=0.
\end{equation}
Due to the fact that the operators $\hat\Pi_i^{\mu_1\mu_2}$  spanning $\rho_{\mathcal G}$ \eqref{rhoInvariant} and operators $\hat\Pi_i^{r_1r_2}$ spanning $\hat\pi_{\mathcal G}^i$ \eqref{Ut1} commute with each other \eqref{PiCommute}. Therefore we have:
\begin{eqnarray}
    \label{def-twirl-gen-proof-2}
    \mathcal T_{\mathcal G}(\rho_{\mathcal G})&=&\int_{\mathcal G}\operatorname{d}g\,f(g)\sum_i\hat\pi_{\mathcal G}^i\rho_{\mathcal G}\,\sum_j\hat\pi_{\mathcal G}^{j\dagger}\nonumber\\
    &=&\rho_{\mathcal G}\int_{\mathcal G}\operatorname{d}g\,f(g)\sum_i\hat\pi_{\mathcal G}^i\,\sum_j\hat\pi_{\mathcal G}^{j\dagger}\nonumber\\
    &=&\rho_{\mathcal G}\int_{\mathcal G}\operatorname{d}g\,f(g)\hat\pi_{\mathcal G}\,\hat\pi_{\mathcal G}^{\dagger}=\rho_{\mathcal G}\int_{\mathcal G}\operatorname{d}g\,f(g)\id=\rho_{\mathcal G}.\nonumber\\
\end{eqnarray}

In the above, we have used the commutativity relation \eqref{RoCommute} and the fact that the integral with respect to the measure $\operatorname{d}g\,f(g)$ is normalized. Note that nowhere in the proof we assumed any restrictions on the choice of the measure $\operatorname{d}g\,f(g)$ apart from the demand that it is normalized. This means that invariant encoding \eqref{rhoInvariant} works for any method of drawing random elements of the symmetry group $\mathcal G$. 

Note that the invariant state \eqref{rhoInvariant} is a sum of invariant states defined within each of the subspaces $P_i$ \eqref{Pi}: 
\begin{equation}
    \label{rho_inv_sum}
    \rho_{\mathcal G}=\sum_i \rho^i_{\textrm{inv}},
\end{equation}
where \begin{equation}
\label{rhoInvI}
    \rho^i_{\textrm{inv}}=\frac{1}{D^i_L}\sum_{\mu_1\mu_2=1}^{D^i_V}\rho^{i}_{\mu_1\mu_2}\hat\Pi_{i}^{\mu_1\mu_2}.
\end{equation}
Whenever for some $i=i_{\textrm{singl}}$ we have $D^{i_{\textrm{singl}}}_L=1$ and at the same time $D^{i_{\textrm{singl}}}_V>1$, then the entire subspace spanned by states $\{\ket{i_{\textrm{singl}}, r, \mu}\}_{\mu}$ is invariant under the action of $\mathcal G$, and therefore states $\rho^{i_{\textrm{singl}}}_{\textrm{inv}}$ are called elements of a Decoherence Free Subspace. Such states are generalizations of well-known singlet states. Whenever both dimensions fulfill $D^i_L>1$ and $D^i_V>1$, then invariant states \eqref{rhoInvI} are called elements of a Decoherence Free Subsystem. The difference is that in the case of Decoherence Free Subspaces the entire $i$-th  subspace $P_i$ \eqref{Pi} is \textit{fixed} under the action of $\mathcal G$, whereas in the case of Decoherence Free Subsystems, the $i$-th subspace $P_i$ is only \textit{invariant} under $\mathcal G$, whereas only specific subsystems in $P_i$ are themselves \textit{fixed}.

The presented construction works for any symmetry group $\mathcal G$ which acts on the state space via unitary representation. This implies that in the case of a compact symmetry group (like e.g. $\textrm{SU}(2)$) the representation space can be both finite and infinite dimensional, whereas in the case of non-compact symmetry groups (like e.g. Lorentz group) the state space is always infinite dimensional. Nevertheless, it may happen that in the context of a non-compact group one applies presented construction to a finite-dimensional subspace of the entire infinite-dimensional state space. This would be the case discussed in further sections of this article, in which we utilize the construction of invariant encoding applied to discrete degrees of freedom of quantum particles, which physically correspond to spin and helicity.

If we relax the assumption of the unitarity of the representation $\hat\pi_{\mathcal G}^i$  the integral $\int_{\mathcal G}\operatorname{d}g\,f(g)\sum_i\hat\pi_{\mathcal G}^i\,\sum_j\hat\pi_{\mathcal G}^{j\dagger}$ is much more difficult to evaluate for an arbitrary measure on the group manifold. In \cite{Markiewicz22} it is shown that in a special case of the measure $\operatorname{d}g$ being a product with respect to the so-called Cartan decomposition of the group $\mathcal G$  and uniform over its compact components it can be easily evaluated and reads $\bigoplus_i\beta_i^f\id_i$, where the coefficients $\beta_i^f$ depend on the integration measure on the non-compact component of the group. In such a case the state $\rho_{\mathcal G}$ is not invariant, but rescaled by the coefficients $\beta_i^f$ within each irreducible subspace $i$.

\section{Quantum particles under Lorentz transformations} \label{sectionII}

In order to discuss the issue of  encoding of quantum information which is invariant with respect to the action of the Lorentz group, we have to describe how general Poincare (inhomogenous Lorentz) transformations affect states of quantum particles. The building blocks of such description are single-particle representations of  the Poincare group $\mathcal P$.  Such representations were classified by Wigner using his famous method of \textit{little groups}, see e.g. \cite{Tung} for detailed presentation. Here we just sketch the most important elements of this classification necessary for discussion of the main tool of this work, namely the multiparticle representations, which are described more thoroughly further in this section.

\subsection{Single-particle representations}
The representation space for single-particle representations of $\mathcal{P}$ is spanned by four-momentum operator  eigenstates, denoted typically as $\{\ket{ p, \sigma}\}$, in which $p$ denotes the eigenvalue of four-momentum, whereas $\sigma$ denotes discrete degree of freedom as spin or helicity. Physical states cannot be represented by sharp-momentum eigenstates, therefore they are in general expressible as:
\begin{equation}
    \label{gen4momState}
    \ket{\psi}=\sum_{\sigma}\int\psi_{\sigma}(p)\ket{p,\sigma}\operatorname{d}\mu(p).
\end{equation}
According to Wigner's classification, there are in general six classes of single-particle representations of $\mathcal P$, three of which have physical meaning. Among these three, one corresponds to vacuum representation, therefore we are left with two non-trivial classes, one corresponding to massive particles, another one to the massless ones. If we denote an arbitrary element of a Poincare group by $(\Lambda, x)$, in which $\Lambda$ is an arbitrary Lorentz transformation, whereas $x$ is a four-vector encoding translation in Minkowski space, these two representations have the following explicit form:
\begin{eqnarray}
\label{singPartWignerIrreps}
    \pi^{(m,s)}_{\mathcal P}(\Lambda, x)\ket{p,\sigma}&=&e^{-ixp}N_{\Lambda,p} \sum_{\sigma'}\operatorname{D}^{(s)}_{\sigma'\sigma}(\Omega_{\Lambda}^p)\ket{\Lambda p,\sigma'},\nonumber\\
     \pi^{(\lambda)}_{\mathcal P}(\Lambda, x)\ket{p,\lambda}&=&e^{-ixp}N_{\Lambda,p} e^{i\lambda\omega_{\Lambda}^p}\ket{\Lambda p,\lambda}.
\end{eqnarray}
In the above formulas $N_{\Lambda,p}$ denotes normalisation factor, $\operatorname{D}^{(s)}_{\sigma'\sigma}$ is a spin-$s$ matrix representation of the special orthogonal group $\textrm{SO}(3)$ \footnote{The easiest way to explicitly express representation matrices $\operatorname{D}^{(s)}(\vec\theta)$, in which $\vec\theta$ is a vector parametrizing an $\textrm{SO}(3)$ rotation, is via matrix exponent of spin-$s$ matrices $J_1^{(s)}, J_2^{(s)}, J_3^{(s)}$:
 $\operatorname{D}^{(s)}(\vec\theta)=\exp\left(-i\vec\theta\cdot\vec J^{(s)}\right)$.}, $\Omega_{\Lambda}^p$ is a concrete set of angles parameterizing an $\textrm{SO}(3)$ rotation, whereas $\omega_{\Lambda}^p$ is an angle parameterizing  $\textrm{SO}(2)$ rotation. These two rotations are typically referred to as \textit{Wigner rotations}, and the corresponding angles as \textit{Wigner rotation angles}.
The dependence of the first representation on the mass $m$ of the particle is hidden in the phase factor  $e^{-ixp}$ and in the Wigner rotation angle $\Omega_{\Lambda}^p$. Indeed $xp=-Et+\vec x\cdot\vec p$, and the energy of the particle has to fulfill the \textit{mass-shell condition} $E=\sqrt{m^2+|\vec p|^2}$.  On the other hand  $\Omega_{\Lambda}^p$ depends implicitly on the Lorentzian ``$\gamma$'' factor equal to $\sqrt{m^2+|\vec p|^2}/m$.
Before we go to a discussion of multipartite representations two remarks are needed at this stage concerning the above representations. 

Firstly, quantum numbers $\sigma$ and $\lambda$ in \eqref{singPartWignerIrreps} play entirely different roles. $\sigma$ in the massive particles representation typically corresponds to the eigenvalue of $\hat S_z$ -- projection of the spin operator onto a distinguished "z" direction in the centre of mass frame, therefore all states $\{\ket{p,\sigma'}\}_{\sigma'=-s}^s$ live within an irreducible spin-$s$ subspace with respect to the Poincare group \footnote{Sometimes one uses a different choice of basis in which $\sigma$ corresponds to massive particle's helicity. We will utilize this approach in Sec. \ref{sec:irreps}.}. On the other hand, the helicity $\lambda$ in the massless particles representation determines different inequivalent irreducible representations of the Poincare group. This inequivalence is explicit due to the presence of $\lambda$  in the exponent of the phase term in the representation formula.

Secondly, the transformation rules for quantum states with a definite value of the discrete variable: spin projection $\sigma$ or helicity $\lambda$ directly depend on the value of the four-momentum of the particle, due to momentum dependence of the Wigner rotation angles $\Omega_{\Lambda}^p$ and $\omega_{\Lambda}^p$. For this reason $\sigma$ and $\lambda$ are called \textit{secondary variables} \cite{Peres03}. This effect significantly hinders the possibility of defining  subspaces invariant with respect to the action of the Lorentz group in the multiparticle case.

\subsection{Multiparticle representations}

\subsubsection{Product (reducible) representations}

In a standard approach to relativistic quantum mechanics, one assumes that multiparticle states transform under product representations of the Poincare group based on single-particle representations \eqref{singPartWignerIrreps}. Let us fix a basis for an $N$-particle space to be $\bigotimes_{\alpha=1}^{N}\ket{p_{\alpha},\sigma_{\alpha}}$ and respectively $\bigotimes_{{\alpha}=1}^{N}\ket{p_{\alpha},\lambda_{\alpha}}$ for massless particles. Then the multiparticle product representations of the Poincare group are defined as:
\begin{eqnarray}
\label{multiPartWignerIrreps}
    &&\pi^{(N,\{m_{\alpha}, s_{\alpha}\})}_{\mathcal P}(\Lambda, x)\bigotimes_{{\alpha}=1}^{N}\ket{p_{\alpha},\sigma_{\alpha}}\nonumber\\
    &&=\bigotimes_{{\alpha}=1}^{N} \pi^{(m_{\alpha},s_{\alpha})}_{\mathcal P}(\Lambda, x)\ket{p_{\alpha},\sigma_{\alpha}},\nonumber\\
     &&\pi^{(N, \{\lambda_{\alpha}\})}_{\mathcal P}(\Lambda, x)\bigotimes_{{\alpha}=1}^{N}\ket{p_{\alpha},\lambda_{\alpha}}\nonumber\\
     &&=\bigotimes_{{\alpha}=1}^{N}\pi^{(\lambda_{\alpha})}_{\mathcal P}(\Lambda, x)\ket{p_{\alpha},\lambda_{\alpha}}.\nonumber\\
\end{eqnarray}
Note that the action of the Poincare group via the above representations is collective, namely on each particle there acts a representation corresponding to the same group element specified by $(\Lambda,x)$. Therefore it can be seen as a relativistic analogue of the collective action of unitary rotations on ordinary $d$-dimensional quantum systems specified by the first formula of \eqref{collU}. The analogy goes further, namely the representations \eqref{multiPartWignerIrreps} are reducible, however their reduction is much more complicated than in the nonrelativistic case of many qu-$d$-its: irreducible representations are specified by the so-called Mandelstam variables, which are eigenvalues of the square of sum of all four-momenta operators and by eigenvalues of the total angular momentum of the particles under consideration \cite{Cox18}. 
We discuss these representations in the next subsection.

\subsubsection{Irreducible representations}
\label{sec:irreps}

Irreducible representations of the Poincare group acting collectively on a multiparticle space are to a certain extent analogous to the ones for collective action of a unitary group on many finite-dimensional subsystems. In both cases irreducible subspaces are labeled by a \textit{total spin}-like variable; the difference is, however, that the decomposition in the case of relativistic particles is complicated by the dependence of Wigner rotations on momentum variables. 
The construction of such representations in its simplest form utilizes the helicity basis for discrete degrees of freedom for both massive and massless particles and is very similar for both cases. There is, however, an important difference, namely in the case of massless particles we have to assume that there exists their centre of mass frame (which implies that their momenta cannot be simultaneously co-linear).

Here we present just a sketch of the construction of these representations needed to discuss their possible application in the relativistically invariant encoding of quantum information (see Section \ref{sec:NonColMoms}), which, to the best of our knowledge, we propose in this work for the first time. In previous works such representations have been extensively utilized in the context of relativistic description of scattering processes of quantum particles, see e.g. \cite{Jiang21, Shu23}.

The construction of irreducible representations of the Poincare group on multiparticle states has two steps, which correspond to fixing values of the two Casimir operators of the Poincare group:
\begin{enumerate}[label=(\roman*)]
    \item change of the product basis $\bigotimes_{\alpha=1}^{N}\ket{p_{\alpha},\lambda_{\alpha}}$ labeled by each particle's momentum and helicity in the \textit{laboratory frame} to the \textit{total momentum basis} $\ket{P,\{p^{\textrm{CM}}_{\beta}\},\{\lambda^{\textrm{CM}}_{\alpha}\}}$ labelled by total four-momentum $P$ and relative momenta $\{p^{\textrm{CM}}_{\beta}\}$ and all particles's helicities $\{\lambda^{\textrm{CM}}_{\alpha}\}$ \textit{in the centre of mass frame}; total momentum basis states are eigenstates of the Casimir operator $\hat P^2=\left(\sum_{{\alpha}=1}^N \hat p_{\alpha}\right)^2$, in which $\hat p_{\alpha}$ is four-momentum operator corresponding to $\alpha$-th particle;
    \item change of the total momentum basis $\ket{P,\{p^{\textrm{CM}}_{\beta}\},\{\lambda^{\textrm{CM}}_{\alpha}\}}$ to the \textit{total spin-$J$ basis} $\ket{P,J,\Sigma_J, \{\lambda^{\textrm{CM}}_{\alpha}\}}$ labeled by total four momentum $P$, total spin $J$,  total helicity $\Sigma_J$ and the helicities in the centre of mass frame, which play the role of \textit{multiplicity index}, namely different sets of helicities $\{\lambda^{\textrm{CM}}_{\alpha}\}$ enumerate distinct subspaces corresponding to equivalent representations of the Poincare group.
\end{enumerate}
Let us now focus on these two steps in more details. The total four-momentum $P$ has components $P=(-\sqrt{\mathfrak{s}+\vec P^2}, \vec P)$, and we have $P^2=\mathfrak{s}$. The value $\mathfrak{s}$, called the \textit{Mandelstam variable}, corresponds to the eigenvalue of the Casimir operator $\hat P^2$:
\begin{equation}
    \hat P^2\ket{P,\{p^{\textrm{CM}}_{\beta}\},\{\lambda^{\textrm{CM}}_{\alpha}\}}=\mathfrak{s}\ket{P,\{p^{\textrm{CM}}_{\beta}\},\{\lambda^{\textrm{CM}}_{\alpha}\}},
\end{equation}
and it represents the total energy of the multiparticle system in the centre of mass frame. Since the value $\mathfrak{s}$ is an eigenvalue of the Casimir operator corresponding to the action of the Poincare group on the multiparticle system under consideration, it is unchanged under arbitrary Poincare transformations and hence labels invariant subspaces under the action of the Poincare group. These subspaces are, however, not irreducible, as full irreducibility is achieved in the second step. 

The second step is much more complicated, as it requires iterated integration over the relative momenta. Let us for further clearness denote irreducible basis vectors by $\ket{\mathfrak{s},J;P,\Sigma_J, \{\lambda^{\textrm{CM}}_{\alpha}\}}$, as we assume that the value of the Mandelstam variable $\mathfrak{s}$ is fixed at this stage. Then we have:
\begin{eqnarray}
    \label{partialWaveInt}
    &&\ket{\mathfrak{s},J;P,\Sigma_J,\{\lambda^{\textrm{CM}}_{\alpha}\}}=\nonumber\\
    &&\mathcal N^{J}\int\operatorname{d}\Omega\left(\{p^{\textrm{CM}}_{\beta}\}\right)\mathcal C^{J,\Sigma_J}_{\{p^{\textrm{CM}}_{\beta}\},\{\lambda^{\textrm{CM}}_{\alpha}\}}\ket{P,\{p^{\textrm{CM}}_{\beta}\},\{\lambda^{\textrm{CM}}_{\alpha}\}}.\nonumber\\
\end{eqnarray}
Formulas of the above type are known as \textit{partial wave decompositions} \cite{Jiang21, Shu23} in the literature and are extensively used in scattering theory. The coefficient $\mathcal N^{J}$ is a normalization factor, the integral $\int\operatorname{d}\Omega\left(\{p^{\textrm{CM}}_{\beta}\}\right)$ is an iterated integral over all relative momenta, see e.g. \cite{Shu23}, Sec. II, and the coefficients $\mathcal C^{J,\Sigma_J}_{\{p^{\textrm{CM}}_{\beta}\},\{\lambda^{\textrm{CM}}_{\alpha}\}}$ are known as \textit{Poincare Clebsch-Gordan} coefficients \cite{Jiang21, Shu23}, in analogy to decomposition of multi-spin states into irreducible subspaces under rotation group. The states $\ket{\mathfrak{s},J;P,\Sigma_J,\{\lambda^{\textrm{CM}}_{\alpha}\}}$ are simultaneous eigenstates of both Casimir operators of the Poincare group acting on a multiparticle state space: 
\begin{itemize}
    \item square of the total momentum operator $\hat P^2=\left(\sum_{\alpha=1}^N \hat p_{\alpha}\right)^2$, with the corresponding eigenvalue $\mathfrak{s}$;
    \item square of the total Pauli-Lubanski operator  $\hat W^2=\left(\sum_{\alpha=1}^N \hat w_{\alpha}\right)^2$, with the corresponding eigenvalue $-\mathfrak{s}J(J+1)$ \footnote{Single-particle Pauli-Lubanski operator is defined as follows in a component-wise form: $\hat w_{\alpha}^{\nu}=\tfrac{1}{2}\epsilon_{\nu\eta\theta\xi}\hat L^{\eta\theta}_{\alpha}\hat p^{\xi}_\alpha$, in which $\epsilon$ is a totally anti-symmetric tensor, $\hat L_{\alpha}$ is a relativistic angular momentum tensor operator of $\alpha$-th particle, and $\hat p_{\alpha}$ is its corresponding four momentum operator.}.
\end{itemize}
The subspaces labeled by pairs of numbers $(\mathfrak{s},J)$ are invariant and irreducible under the action of the Poincare group:
\begin{eqnarray}
    \label{PoincareIrreps}
     &&\pi^{(\mathfrak{s},J)}_{\mathcal P}(\Lambda, x)\ket{\mathfrak{s},J;P,\Sigma_J,\{\lambda^{\textrm{CM}}_{\alpha}\}}=\nonumber\\
     &&e^{-ixP}N_{\Lambda,P} \sum_{\Sigma_J'}\operatorname{D}^{(J)}_{\Sigma_J'\Sigma_J}(\Omega_{\Lambda}^P)\ket{\mathfrak{s},J;\Lambda P,\Sigma_J',\{\lambda^{\textrm{CM}}_{\alpha}\}}.\nonumber\\
\end{eqnarray}

From the above formula it is clear why $\{\lambda^{\textrm{CM}}_{\alpha}\}$ are degeneracy indices: all states corresponding to the same set of values of helicities in the centre of mass frame transform exactly in the same way under the action of the Poincare group.

Note an interesting property arising from the above formula in the case of massless particles: irreducible states of massless particles with total momentum $P$ transform under Poincare transformations as states of massive particles of spin-$J$ and mass squared $M^2=\mathfrak s$ \cite{Shu23}.  In scattering theory one often treats the product basis states $\bigotimes_{\alpha=1}^{N}\ket{p_{\alpha},\lambda_{\alpha}}$ as representing in-states, whereas the irreducible states  $\ket{\mathfrak{s},J;P,\Sigma_J,\{\lambda^{\textrm{CM}}_{\alpha}\}}$ as representing  out-states. Then the Poincare Clebsch-Gordan coefficients $\mathcal C^{J,\Sigma_J}_{\{p^{\textrm{CM}}_{\beta}\},\{\lambda^{\textrm{CM}}_{\alpha}\}}$ play the role of scattering amplitudes, and the entire process is understood as a process in which a product  in-state of $N$ massless particles is transformed into the out-state of a massive spin-$J$ particle \cite{Arkani21, Shu23}.
In the simplest case of two particles with helicities in the centre of mass frame $\lambda^{\textrm{CM}}_1, \lambda^{\textrm{CM}}_2$ in a product in-state the  coefficients $\mathcal C^{J,\Sigma_J}_{\{p^{\textrm{CM}}_{\beta}\},\{\lambda^{\textrm{CM}}_{\alpha}\}}$ read \cite{Shu23}:
\begin{equation}
    \mathcal C^{J,\Sigma_J}_{\{p^{\textrm{CM}}\},\{\lambda^{\textrm{CM}}_1,\lambda^{\textrm{CM}}_2 \}}=\operatorname{D}^{(J)*}_{\Sigma_J, \lambda^{\textrm{CM}}_1-\lambda^{\textrm{CM}}_2}\left(\Omega\left(p^{\textrm{CM}}\right)\right),
\end{equation}
in which $p^{\textrm{CM}}$ is a relative momentum of the particles in the centre of mass frame and the star $^*$ denotes complex conjugation.  In this case, the partial wave decomposition formula \eqref{partialWaveInt} reads:
\begin{eqnarray}
    &&\ket{\mathfrak{s},J;P,\Sigma_J,\{\lambda^{\textrm{CM}}_1, \lambda^{\textrm{CM}}_2\}}=\nonumber\\
    &&\sqrt{\frac{2J+1}{4\pi}}\int\operatorname{d}\Omega\left(p^{\textrm{CM}}\right)\operatorname{D}^{(J)*}_{\Sigma_J, \lambda^{\textrm{CM}}_1-\lambda^{\textrm{CM}}_2}\left(\Omega\left(p^{\textrm{CM}}\right)\right)\nonumber\\
    &&\ket{P,p^{\textrm{CM}},\{\lambda^{\textrm{CM}}_1, \lambda^{\textrm{CM}}_2\}}.
\end{eqnarray}

\subsubsection{Pairwise helicity representations}
The formalism  presented above is highly general and sufficient in most circumstances. However,  as pointed out in \cite{Zwanziger72} product basis states $\bigotimes_{{\alpha}=1}^{N}\ket{p_{\alpha},\sigma_{\alpha}}$ are not enough to describe at least the scattering of electric and magnetic charges. Let us recall a more general construction of multiparticle representations of the Poincar\'e group. The necessary extension turns out to be an inclusion of pair-wise little groups of transformations under which a given \textit{pair of momenta} remains invariant \cite{Csaki21}. In general, any laboratory frame can always be transformed to the centre of momentum frame for a given pair of two  particles in such a way that both particles move along the \textit{z}-axis. Then the Lorentz subgroup that leaves the pair of momenta unchanged is isomorphic to $U(1)$ and corresponds to rotations around \textit{z}-axis and therefore serves as a pair-wise little group. Note that there is no higher particle-number non-trivial little group as there is no Lorentz transformation, which would in general keep the triple or more of momenta invariant. Existence of the pair-wise little group results in the appearance of an additional quantum number $q_{\alpha\alpha'}$ called pair-wise helicity. Taking this into consideration, the general form of the multi-particle state is no longer product with respect to single-particle states and is given by:
\begin{eqnarray}
    &&\ket{p^N;\sigma^N(\lambda^N);\lbrace q_{{\alpha}{\alpha'}}\rbrace}\equiv\nonumber\\  &&\ket{p_1,\dots,p_N;\sigma_1(\lambda_1),\dots,\sigma_N (\lambda_N);q_{12},q_{13},\dots,q_{(N-1)N}}.\nonumber\\
\end{eqnarray}
The notation $\sigma^N(\lambda^N)$ and  $\sigma_{\alpha}(\lambda_{\alpha})$  indicates that the above formula works for both massive and massless particles, the states of which are labelled by respectively spin projection and helicity.
Quantum Poincar\'e transformations act on these generalized states as follows:
\begin{itemize}
    \item on the momentum and spin part of the state they act in the same manner as for the product basis \eqref{multiPartWignerIrreps};
    \item on the  pair-wise helicity part of the state they act in the same way as on the ordinary helicity of massless particles,  resulting in a phase gain $\phi_{\Lambda,{\alpha}{\alpha'}}$:
\end{itemize}
\begin{widetext}
\begin{eqnarray}\label{eq:pair_wise_transf}
    &&\pi^{(N,\{m_{\alpha}, s_{\alpha},q_{{\alpha}{\alpha'}}\})}_{\mathcal P}(\Lambda, x)\ket{p^N;\sigma^N;\lbrace q_{{\alpha}{\alpha'}}\rbrace}\equiv\nonumber\nonumber\\&&=\prod_{{\beta}>{\beta'}}e^{iq_{{\beta}{\beta'}}\phi_{\Lambda,{\beta}{\beta'}}}\prod_\alpha e^{-ixp_{\alpha}}N_{\Lambda,p_{\alpha}} \sum_{\sigma'_{\alpha}}\operatorname{D}^{(s_{\alpha})}_{\sigma'_{\alpha}\sigma_{\alpha}}(\Omega_{\Lambda}^{p_{\alpha}})\ket{\Lambda p_1,\dots,\Lambda p_N;\sigma'_1,\dots,\sigma'_N ;q_{12},q_{13},\dots,q_{(N-1)N}},\nonumber\\
     &&\pi^{(N, \{\lambda_{\alpha},q_{{\alpha}{\alpha'}}\})}_{\mathcal P}(\Lambda, x)\ket{p^N;\lambda^N;\lbrace q_{{\alpha}{\alpha'}}\rbrace}\equiv\nonumber\nonumber\\&&=\prod_{{\beta}>{\beta'}}e^{iq_{{\beta}{\beta'}}\phi_{\Lambda,{\beta}{\beta'}}}\prod_\alpha N_{\Lambda,p_{\alpha}}  e^{-ixp_{\alpha}}e^{i\lambda_{\alpha}\omega_{\Lambda}^{p_{\alpha}}}\ket{\Lambda p_1,\dots,\Lambda p_N;\lambda_1,\dots,\lambda_N ;q_{12},q_{13},\dots,q_{(N-1)N}}.\nonumber\\
\end{eqnarray}
\end{widetext}
In most physical situations one has $q_{{\alpha}{\alpha'}}=0$ and therefore this representation is equivalent to the one based on the product basis states. The scenario in which it is known that $q_{{\alpha}{\alpha'}}\neq 0$ is the process of scattering of the electric-magnetic charges (dyons). In such scenario, pair-wise helicity is given by:
\begin{equation}
    q_{{\alpha}{\alpha'}}= e_{\alpha}g_{\alpha'}-e_{\alpha'}g_{\alpha},\label{eq:pair_wise_helicity}
\end{equation}
where $e_{\alpha}$ stands for electric charge and $g_{\alpha}$ for magnetic charge of the ${\alpha}$-th particle. The appearance of nonzero $q_{{\alpha}{\alpha'}}$ is a consequence of the fact that the angular momentum stored in electromagnetic field of electric-magnetic charges gives additional contribution to the total angular momentum of the system, which is independent of the distance between two-particles when electric and magnetic charges are concerned.

\section{Relativistically invariant encoding of quantum information}

\subsection{Equal momentum encodings} \label{sectionIII_col}

In this group of encodings, we follow the idea introduced in \cite{Bartlett05}, in which one encodes quantum information using states of $N$ distinguishable (due to position degree of freedom) quantum particles with almost sharp and equal momenta. Fulfilling these two conditions simultaneously seems  contradictory (sharp momentum implies spatial delocalization), however, as shown in \cite{Bartlett05} one can come close to fulfilling them. The assumptions of using states of particles with equal and sharp momenta are motivated by an attempt to allow for invariant encoding with spin/helicity degrees of freedom at the same time avoiding decoherence  due to entaglement between momentum and spin/helicity, which arises due to Lorentz transformations whenever momentum wavepackets are used instead of sharp momenta. In this section we discuss three cases, two already known: for massive and massless particles, and third one, for massive particles with pairwise helicity, proposed in this article for the first time in the context of hypothetical electric-magnetic charges (dyons).

\subsubsection{Massive particles}

A typical scenario for the encoding of quantum information into states of many massive particles, which is invariant with respect to the action of a Lorentz group $\mathcal L$, goes as follows \cite{Bartlett05}. First, one prepares a $N$-partite state of spin-$\tfrac{1}{2}$ particles in the form of a one-dimensional lattice with a constant spatial separation such that all the particles have a common almost sharp momentum $p$ and at the same time their single particle overlaps are arbitrarily small such that the particles are effectively distinguishable (localized). Then the state of such particles can be represented as:
\begin{equation}
    \label{psiN}
    \ket{\psi}^{p,m>0}=\sum_{{\alpha}_1,\ldots,{\alpha}_N=\pm 1}c_{{\alpha}_1\ldots {\alpha}_N}\ket{p,\sigma_{{\alpha}_1}}\ldots\ket{p,\sigma_{{\alpha}_N}}, 
\end{equation}
in which $\sigma_{\pm 1}$ denotes up/down state of spin-$\tfrac{1}{2}$ projection onto local particle's "z" axis.
The Lorentz transformation acting on each single particle acts via the first (massive) representation in \eqref{singPartWignerIrreps} with $x=0$ and $s=\tfrac{1}{2}$:
\begin{equation}
    \label{LorentzSingPart}
    \pi_{\mathcal L}^{(\frac{1}{2})}(\Lambda)\ket{p,\sigma}=N_{\Lambda,p}\operatorname{D}^{(\frac{1}{2})}(\Omega_{\Lambda}^p)\ket{\Lambda p,\sigma},
\end{equation}
in which $\operatorname{D}^{(\frac{1}{2})}$ is just a spin-$\tfrac{1}{2}$ representation of $\textrm{SO}(3)$. We skipped the massive case indicator $(m)$ in the notation for this representation as it is assumed that all particles have equal mass, since otherwise the Wigner rotation angle would be different for each particle under the same Lorentz transformation $\Lambda$. If we denote the state of the entire system by $\rho^{p,m>0}=\ket{\psi}^{p,m>0}\!\bra{\psi}$ then the twirling map with respect to the collective action of a Lorentz group via representation \eqref{LorentzSingPart} takes the form:
\begin{equation}
    \label{LorentzTwirlMass}
    \mathcal T_{\mathcal L}(\rho^{p,m>0})=\int_{\mathcal L}\operatorname{d}\!\Lambda f(\Lambda)\left[\pi_{\mathcal L}^{(\frac{1}{2})}(\Lambda)\right]^{\otimes N}\rho^{p,m>0} \left[\pi_{\mathcal L}^{(\frac{1}{2})}(\Lambda)^{\dagger}\right]^{\otimes N}.
\end{equation}
Note that the above integral is performed over non-compact group manifold of the Lorentz group, therefore $\operatorname{d}\!\Lambda f(\Lambda)$ must be chosen in a way which assures convergence of the integral.
In fact, we demand that the above integral is normalised, $\int_{\mathcal L}\operatorname{d}\!\Lambda f(\Lambda)=1$.
This remark remains valid throughout the rest of this work whenever we consider twirling operations over Lorentz group.
Let us now take a closer look at this map. For clarity of notation, let us denote the initial state as follows:
\begin{equation}
    \ket{\psi}^{p,m>0}=\sum_{k}c_{n^{(k)}}\bigotimes_{\alpha}\ket{p, \sigma_{n^{(k)}_{\alpha}}},
\end{equation}
and the corresponding density matrix as:
\begin{equation}
    \rho^{p,m>0}=\sum_{k,l}c_{n^{(k)}}c_{n^{(l)}}^*\bigotimes_{{\alpha},{\alpha'}}\ket{p, \sigma_{n^{(k)}_{\alpha}}}\bra{\sigma_{n^{(l)}_{\alpha'}},p},
\end{equation}
where $n^{(k)}$ is the $N$-tuple multi-index, and $n^{(k)}_{\alpha}=\pm 1$ is its ${\alpha}$-th element.
Then by inserting \eqref{LorentzSingPart} into \eqref{LorentzTwirlMass} we obtain:
\begin{eqnarray}
\label{spinTwirl}
     &&\mathcal T_{\mathcal L}(\rho^{p,m>0})=\int_{\mathcal L}\operatorname{d}\!\Lambda f(\Lambda)N_{\Lambda,p}^2\sum_{k,l}c_{n^{(k)}}c_{n^{(l)}}^*\nonumber\\
     &&\left[\operatorname{D}^{(\frac{1}{2})}(\Omega_{\Lambda}^p)\right]^{\otimes N}\bigotimes_{{\alpha},{\alpha'}}\ket{\Lambda p,\sigma_{n^{(k)}_{\alpha}}}\bra{\sigma_{n^{(l)}_{\alpha'}},\Lambda p}\left[\operatorname{D}^{(\frac{1}{2})}(\Omega_{\Lambda}^p)^{\dagger}\right]^{\otimes N}.\nonumber\\
\end{eqnarray}
Now because all the components in the tensor product share the same value of $p$, the action of the Wigner rotation $\operatorname{D}^{(\frac{1}{2})}(\Omega_{\Lambda}^p)$ on spin degrees of freedom is collective:
\begin{eqnarray}
\label{spinTwirl1}
     &&\mathcal T_{\mathcal L}(\rho^{p,m>0})=\int_{\mathcal L}\operatorname{d}\!\Lambda f(\Lambda)N_{\Lambda,p}^2\left(\ket{\Lambda p}\bra{\Lambda p}\right)^{\otimes N}\otimes\sum_{k,l}c_{n^{(k)}}c_{n^{(l)}}^*\nonumber\\
     &&\left[\operatorname{D}^{(\frac{1}{2})}(\Omega_{\Lambda}^p)\right]^{\otimes N}\bigotimes_{{\alpha},{\alpha'}}\ket{\sigma_{n^{(k)}_{\alpha}}}\bra{\sigma_{n^{(l)}_{\alpha'}}}\left[\operatorname{D}^{(\frac{1}{2})}(\Omega_{\Lambda}^p)^{\dagger}\right]^{\otimes N},\nonumber\\
\end{eqnarray}
and therefore reducible in a standard way discussed in Sec. II.B. By changing the product spin basis to the basis  irreducible with respect to collective rotations:
\begin{eqnarray}
\label{spinTwirl2}
     &&\mathcal T_{\mathcal L}(\rho^{p,m>0})=\int_{\mathcal L}\operatorname{d}\!\Lambda f(\Lambda)N_{\Lambda,p}^2\left(\ket{\Lambda p}\bra{\Lambda p}\right)^{\otimes N}\otimes\sum_{J,r,r',\mu,\mu'}\nonumber\\
     &&c^J_{r,\mu}c^{J*}_{r',\mu'}
     \operatorname{D}^{(J)}(\Omega_{\Lambda}^p)\ket{J,r,\mu}\bra{\mu',r',J}\operatorname{D}^{(J)}(\Omega_{\Lambda}^p)^{\dagger},\nonumber\\
\end{eqnarray}
in which the irreducible representations of the rotation group are labelled by the total spin $J$,
one can utilize invariant encoding \eqref{rhoInvariant} discussed in Section \ref{sec:genInvState}. Namely let us assume an initial state in the following form:
\begin{equation}
    \label{inStateMassive}
    \rho^{p,m>0}_{\mathcal L}=\left(\ket{p}\bra{p}\right)^{\otimes N}\otimes \rho^{\operatorname{sp}}_{\mathcal L},
\end{equation}
in which $\rho^{\operatorname{sp}}_{\mathcal L}$ is exactly the Lorentz-invariant state \eqref{rhoInvariant} defined on spin degrees of freedom:
\begin{equation}
    \label{rhoInvariantMassive}
    \rho^{\operatorname{sp}}_{\mathcal L}=\sum_{J}\frac{1}{D^J_L}\sum_{\mu_1\mu_2=1}^{D^J_V}\rho^{J}_{\mu_1\mu_2}\hat\Pi_{J}^{\mu_1\mu_2}.
\end{equation}
The operators $\hat\Pi_{J}^{\mu_1\mu_2}$ are defined according to \eqref{PiBasis}:
\begin{equation}
    \hat\Pi^{\mu_1\mu_2}_J=\sum_{r=1}^{D^J_L}\ket{J,r,\mu_1}\bra{\mu_2,r,J}.
\end{equation}
Now since $\rho^{\operatorname{sp}}_{\mathcal L}$ as Lorentz-invariant state commutes with each ${\operatorname{D}^{(J)}}$, and  ${\operatorname{D}^{(J)}}$ is unitary, we have:
\begin{eqnarray}
\label{spinTwirl3}
     \mathcal T_{\mathcal L}(\rho^{p,m>0}_{\mathcal L})&=&\int_{\mathcal L}\operatorname{d}\!\Lambda f(\Lambda)N_{\Lambda,p}^2\left(\ket{\Lambda p}\bra{\Lambda p}\right)^{\otimes N}\otimes\sum_J\nonumber\\
     &&\operatorname{D}^{(J)}(\Omega_{\Lambda}^p)\rho^{\operatorname{sp}}_{\mathcal L}\operatorname{D}^{(J)}(\Omega_{\Lambda}^p)^{\dagger}\nonumber\\
     &=&\int_{\mathcal L}\operatorname{d}\!\Lambda f(\Lambda)N_{\Lambda,p}^2\left(\ket{\Lambda p}\bra{\Lambda p}\right)^{\otimes N}\otimes\rho^{\operatorname{sp}}_{\mathcal L}\nonumber\\
     &&\sum_J \operatorname{D}^{(J)}(\Omega_{\Lambda}^p)\operatorname{D}^{(J)}(\Omega_{\Lambda}^p)^{\dagger}\nonumber\\
    &=&\left(\int_{\mathcal L}\operatorname{d}\!\Lambda f(\Lambda)N_{\Lambda,p}^2\left(\ket{\Lambda p}\bra{\Lambda p}\right)^{\otimes N}\right)\otimes\rho^{\operatorname{sp}}_{\mathcal L}.\nonumber\\
\end{eqnarray}
This shows that the spin component $\rho^{\operatorname{sp}}_{\mathcal L}$ of the state $\rho^{p,m>0}_{\mathcal L}$ \eqref{inStateMassive} is invariant with respect to the action of the Lorentz group. Note the crucial property of such encoding, namely that it works irrespective of the way of sampling the Wigner rotation angles $\Omega_{\Lambda}^p$. This is very important, since typically such sampling would not be done according to the Haar measure on the rotation group. This completes the proofs of invariance provided by \cite{Bartlett05} and in \cite{Bartlett07}, Sec. II C, which both assume uniform  sampling of the rotation angles.  

\subsubsection{Massless particles}

Let us now discuss the case of invariant encoding of quantum information in helicity degree of freedom for massless particles. Further on we will utilize the same idea for encoding based on pairwise helicity in the case of massive particles. The initial state encoded in momentum-helicity basis is now specified by:
\begin{equation}
    \ket{\psi}^{p,m=0}=\sum_{k}c_{n^{(k)}}\bigotimes_{{\alpha}=1}^N\ket{p, \lambda_{n^{(k)}_{\alpha}}},
\end{equation}
where $n^{(k)}$ is the $N$-tuple multi-index, and $n^{(k)}_{\alpha}=\pm 1$ is its ${\alpha}$-th element. $\lambda_{\pm 1}$ represents left/right-handed helicity of a particle. The corresponding density matrix reads:
\begin{equation}
    \rho^{p,m=0}=\sum_{k,l}c_{n^{(k)}}c_{n^{(l)}}^*\bigotimes_{{\alpha},{\alpha'}=1}^N\ket{p, \lambda_{n^{(k)}_{\alpha}}}\bra{\lambda_{n^{(l)}_{\alpha'}},p}.
\end{equation}
Lorentz transformation $\Lambda$ then acts on each particle via:
\begin{equation}
    \pi^{\lambda}_{\mathcal L}(\Lambda)\ket{p,\lambda}=N_{\Lambda,p} e^{i\lambda\omega_{\Lambda}^p}\ket{\Lambda p,\lambda},
\end{equation}
therefore the twirling map reads:
\begin{equation}
    \label{LorentzTwirlMassless}
    \mathcal T_{\mathcal L}(\rho^{p,m=0})=\int_{\mathcal L}\operatorname{d}\!\Lambda f(\Lambda)\left[\pi_{\mathcal L}^{\lambda}(\Lambda)\right]^{\otimes N}\rho^{p,m=0} \left[\pi_{\mathcal L}^{\lambda}(\Lambda)^{\dagger}\right]^{\otimes N}.
\end{equation}
In detail, the above map reads:
\begin{eqnarray}
\label{masslessTwirl}
     \mathcal T_{\mathcal L}(\rho^{p,m=0})&=&\int_{\mathcal L}\operatorname{d}\!\Lambda f(\Lambda)N_{\Lambda,p}^2\left(\ket{\Lambda p}\bra{\Lambda p}\right)^{\otimes N}\otimes\sum_{k,l}c_{n^{(k)}}c_{n^{(l)}}^*\nonumber\\
     &&\bigotimes_{{\alpha},{\alpha'}}e^{i\lambda_{n^{(k)}_{\alpha}}\omega_{\Lambda}^p}\ket{\lambda_{n^{(k)}_{\alpha}}}\bra{\lambda_{n^{(l)}_{\alpha'}}}e^{-i\lambda_{n^{(l)}_{\alpha'}}\omega_{\Lambda}^p}\nonumber\\
     &=&\int_{\mathcal L}\operatorname{d}\!\Lambda f(\Lambda)N_{\Lambda,p}^2\left(\ket{\Lambda p}\bra{\Lambda p}\right)^{\otimes N}\otimes\sum_{k,l}c_{n^{(k)}}c_{n^{(l)}}^*\nonumber\\
     &&e^{i\left(\sum_{\alpha}\lambda_{n^{(k)}_{\alpha}}-\sum_{\alpha'}\lambda_{n^{(l)}_{\alpha'}}\right)\omega_{\Lambda}^p}\bigotimes_{{\alpha},{\alpha'}}\ket{\lambda_{n^{(k)}_{\alpha}}}\bra{\lambda_{n^{(l)}_{\alpha'}}}.\nonumber\\
\end{eqnarray}
Now note that if we restrict to states for which the sum of helicities corresponding to each term is constant:
\begin{equation}
    \label{constSumHel}
\forall_k\,\, \sum_{\alpha=1}^N\lambda_{n^{(k)}_{\alpha}}=h,
\end{equation}
the phase factor in the fourth row of  \eqref{masslessTwirl} disappears and  then the helicity part of the states corresponding to such a subspace is untouched by the action of the Lorentz group. Therefore, the invariant encoding is in this case realized by helicity states with constant sum of helicities.

Let us now formally place this result within the general scheme of Sec. \ref{sec:genInvState}.
For simplicity, let us assume that the helicity takes just two values $\pm 1$ (this is the only physically relevant case, however, formal generalization to more values acquired by the helicity is straightforward).
Since we assume that the common four-momentum $p$ is fixed, the irreducible subspaces under the action of the Lorentz group are solely determined by irreducible representations of the Wigner's little group for massless particles, namely the group $\operatorname{U}(1)$.
The irreducible subspaces are labeled by the number $h$, which takes values from among numbers: 
$h=-N,-N+2,\ldots,N-2,N.$ The decomposition \eqref{SchurBasis} reads in this case:
 \begin{eqnarray}
    \label{SchurBasisU1}
L^h_1:\,\,&&\ket{h,1,1}=\bigotimes_{{\alpha}=1}^N\ket{\lambda_{n^{(1)}_{\alpha}}}\nonumber\\
L^h_2:\,\,&&\ket{h,1,2}=\bigotimes_{{\alpha}=1}^N\ket{\lambda_{n^{(2)}_{\alpha}}}\nonumber\\
     && \ldots\nonumber\\
L^h_{D^h_V}:\,\,&&\ket{h,1,D^h_V}=\bigotimes_{{\alpha}=1}^N\ket{\lambda_{n^{\left(D^h_V\right)}_{\alpha}}},
    \end{eqnarray}
    where all the helicity states fulfill condition \eqref{constSumHel}.
All equivalent irreducible representations (corresponding to the rows of the above array) are one-dimensional, since the Wigner's little group $\operatorname{U}(1)$  is Abelian.  The multiplicity of each irreducible subspace, namely the number of rows in the above array, is equal to the number of permutations of a set consisting of  $(N+h)/2$ ones and $(N-h)/2$ minus ones (note that both these numbers are integers by construction), namely:
\begin{equation}
    \label{dimNoiselessMassless}
    D^h_V=\frac{N!}{\left(\frac{N+h}{2}\right)!\left(\frac{N-h}{2}\right)!}.
\end{equation}
In order to explicitly construct an invariant encoding, we take the following initial state:
\begin{equation}
    \label{inStateMassless}
    \rho^{p,m=0}_{\mathcal L}=\left(\ket{p}\bra{p}\right)^{\otimes N}\otimes \rho^{\operatorname{hel}}_{\mathcal L},
\end{equation}
in which $\rho^{\operatorname{hel}}_{\mathcal L}$ is the Lorentz-invariant state \eqref{rhoInvariant} defined on helicity degrees of freedom:
\begin{equation}
    \label{rhoInvariantMassless}
    \rho^{\operatorname{hel}}_{\mathcal L}=\sum_{h}\sum_{\mu_1\mu_2=1}^{D^h_V}\rho^{h}_{\mu_1\mu_2}\hat\Pi_{h}^{\mu_1\mu_2}.
\end{equation}
Operators $\hat\Pi_{h}^{\mu_1\mu_2}$ are, in this case, simply defined by:
\begin{equation}
    \hat\Pi^{\mu_1\mu_2}_h=\ket{h,1,\mu_1}\bra{\mu_2,1,h}.
\end{equation}
Then we have:
\begin{equation}
    \label{rhoInvariantMassless1}
     \mathcal T_{\mathcal L}(\rho^{p,m=0}_{\mathcal L})=\left(\int_{\mathcal L}\operatorname{d}\!\Lambda f(\Lambda)N_{\Lambda,p}^2\left(\ket{\Lambda p}\bra{\Lambda p}\right)^{\otimes N}\right)\otimes \rho^{\operatorname{hel}}_{\mathcal L}.
\end{equation}
The invariant encoding is in this case realized solely within Decoherence Free \textit{Subspaces} labeled by the number $h$, each of dimension \eqref{dimNoiselessMassless}.

\subsubsection{Hypothetical electric-magnetic charges}
Let us also consider a different possibility of relativistically invariant encoding of information, which has not been explored before.  As mentioned in previous sections, the state space for the hypothetical electric-magnetic charges is more intricate, giving access to additional quantum numbers i.e. pair-wise helicity $q_{{\alpha}{\alpha'}}$. What is more, the number of these quantum numbers grows quadratically as $N(N-1)/2$ with a number of particles and therefore gives hope for the possibility of denser coding of information than in the typical case of spin-$\tfrac{1}{2}$ particles.

\begin{figure}
    \centering
    \includegraphics[width=\linewidth]{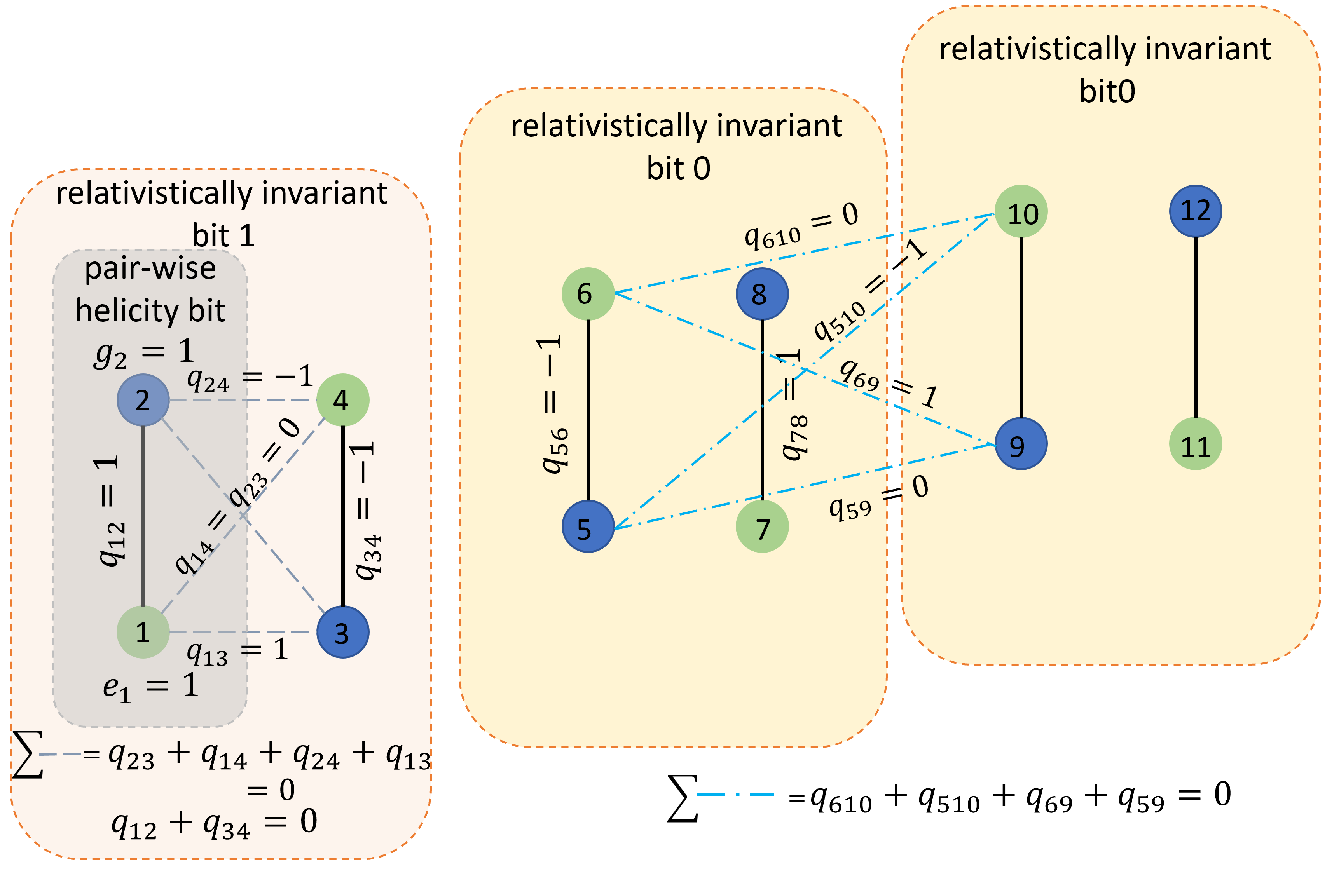}
    \caption{Schematic representation of bits (qubits) encoded in pair-wise helicity and relativistically invariant encoding of bits (qubits). The pair of particles denoted as $1$ and $2$ with charges  ($e_1=1,g_1=0$) and  ($e_2=0,g_2=1$) respectively encodes one of the logical values while pair of particles $3$ and $4$ with the charges reversed encodes the second logical value in pair-wise helicity. The cell created by considering these two pairs of particles has sum of pair-wise helicities equal to 0. The same holds for the cell created with particles $5,6,7,8$ as configuration in this cell is just a permutation of the indices in pairs which can only result in sign change of pair-wise helicities based on \eqref{eq:pair_wise_helicity}. This allows for relativistically invariant encoding of bits (qubits). Addition of further cells preserves the property of the system that the sum of pair-wise helicities is equal to 0. This is implied by the fact that addition of an extra pair in the system is associated with appearance of pair-wise helicities between previous pairs and the new ones in the form presented by dashed lines or dot-dashed lines or with the indices permuted inside pairs. In all those situations, the sum of introduced pair-wise helicities is equal to 0.}
    \label{fig:1}
\end{figure}
We start by introducing the simplest encoding of  bits in terms of pair-wise helicity. Consider the two particles (see Fig. \ref{fig:1}) where one of them is a magnetic monopole ($e_{\alpha}=0,g_{\alpha}=1$) while the second one is a particle with positive electric charge ($e_{\alpha}=1,g_{\alpha}=0$). Assume also that they have masses $m_g$ and $m_e$ respectively, and that all the particles have the same spin $s$. Then if the first particle is a magnetic monopole the pair-wise helicity $q_{12}$ \eqref{eq:pair_wise_helicity} is equal to $-1$ while if the second one is magnetic monopole then we have $q_{12}=1$. Therefore, the configuration of these particles simply encodes a bit. Now, one encodes qubit as a superposition:
\begin{equation}
    \zeta\ket{p_1,p_2,\sigma_1,\sigma_2;1}+\zeta' \ket{p_1,p_2,\sigma_1,\sigma_2;-1},
\end{equation}
where $\zeta,\zeta'$ stands for complex amplitudes. This superposition corresponds to the \textit{which-way} superposition i.e. which path the respective particles have taken.  Here appears the reason why one cannot use pair-wise helicity $q_{{\alpha}{\alpha'}}=0$ as a part of the encoding of qubits, i.e. superselection rules \cite{SSR2}. Note that $q_{{\alpha}{\alpha'}}=0$ appears when we have two particles of the same type (the same charge) while $q_{{\alpha}{\alpha'}}\neq0$ appears for particles with different charges. Therefore, in general, encoding qubits utilizing $q_{{\alpha}{\alpha'}}=0$ would require superpositions of states with different total charges. However, the superselection rules for conserved charges \cite{SSR3,SSR4,SSR1} do not allow such superpositions. Then the state could be only a classical mixture of states with different charges, and thus such a state is unable to encode quantum information.  However, one can superpose the paths of two particles with different charges as scattering of particles does not determine the output paths of the particles and the resulting state is a superposition of states with the same total charge and thus allowed by superselection rules. Note that in the same time bits or qubits can be encoded in the spin degrees of freedom of those particles, as they are independent from pair-wise helicity. 

Now to encode multiple bits or qubits one can build two parallel chains of particles consisting of pairs of  a magnetic monopole and positively charged particle and then use the quantum numbers corresponding to the given pair as above. In such encoding not all pair-wise helicities are used to encode information as for example the pair-wise helicities $q_{12},q_{34}$ fully determine the $q_{13},q_{23},q_{14},q_{24}$. Still, one gets additional qubit for each pair added to the chain and one could consider more complicated encodings to utilize higher number of pair-wise helicities.

Let us now consider the action of twirling with respect to the Lorentz group on the states of the form:
\begin{equation}
    \ket{\psi}^{p,q,m_{e,g}>0}=\sum_{k}c_{n^{(k)}}\ket{p^N; \sigma^N_{n^{(k)}};\lbrace q_{{\alpha}{\alpha'}}^k\rbrace},
\end{equation}
and its corresponding density matrix $\rho^{p,q,m_{e,g}>0}$, where $\lbrace q_{{\alpha}{\alpha'}}^k\rbrace$ stands for configurations of pair-wise helicities for some state indexed by $k$. Here we also assume that the momenta of all particles \textit{of given species} are equal. Let us also alternatively write the basis vectors by the introduction of the virtual tensor product:
\begin{multline}
    \ket{p^N, \sigma^N_{n^{(k)}},\lbrace q_{{\alpha}{\alpha'}}^k\rbrace}=\ket{p^N; \sigma^N_{n^{(k)}}}\otimes\ket{(p^N),\lbrace q_{{\alpha}{\alpha'}}^k\rbrace}\\
    \equiv\bigotimes_{\alpha=1}^N\ket{p_\alpha, \sigma_{n^{(k)}_{\alpha}}}\bigotimes_{\beta>\beta'}^N\ket{(p_{\beta},p_{\beta'}), q^k_{\beta\beta'}}.
\end{multline}
This is to mark that based on \eqref{eq:pair_wise_transf} the basis states transform under Poincar\'e group separately in the spin part and the pair-wise helicity part with only connection being the momenta which in our case come as  parameters parametrizing phase and Wigner's rotation angles. Note that we have left momenta as degrees of freedom which undergoe transformations only in the spin part, while on the  pair-wise helicity part they are left only as parameters. Accordingly, the Poincar\'e transformation factorizes into the part acting on the spin part and pair-wise helicity part (see \eqref{eq:pair_wise_transf}):
\begin{multline}
\pi^{(N, \{m_\alpha,s_{\alpha},q_{{\alpha}{\alpha'}}\})}_{\mathcal P}(\Lambda, x)=\\\left[\pi_{\mathcal P}^{(m,s)}(\Lambda,x)\right]^{\otimes N}\otimes\left[\pi_{\mathcal L}^{q}(\Lambda)\right]^{\otimes N(N-1)/2},
\end{multline}
where $\pi_{\mathcal L}^{q}(\Lambda)$ acts on single pair-wise helicity state resulting in the phase gain parameterized by the momenta which is induced by the Lorentz transformation i.e. 
\begin{multline}
    \pi_{\mathcal L}^{q}(\Lambda)\ket{(p_\alpha,p_{\alpha'}), q^k_{\alpha\alpha'}}=\\e^{iq^k_{\alpha\alpha'}\phi_{\Lambda,\alpha\alpha'}(p_\alpha,p_{\alpha'})}\ket{(p_\alpha,p_{\alpha'}), q^k_{\alpha\alpha'}}.
\end{multline}
Writing down the collective twirling operation with respect to the Lorentz group  on the density matrix $\rho^{p,q,m_{e,g}>0}$ we get:
\begin{widetext}
\begin{equation}
    \mathcal T_{\mathcal L}(\rho^{p,q,m_{e,g}>0})=\int_{\mathcal L}\operatorname{d}\!\Lambda f(\Lambda)\left[\pi_{\mathcal L}^{(m,s)}(\Lambda)\right]^{\otimes N}\left[\pi_{\mathcal L}^{q}(\Lambda)\right]^{\otimes N(N-1)/2}\rho^{p,q,m_{e,g}>0} \left[\pi_{\mathcal L}^{q}(\Lambda)^{\dagger}\right]^{\otimes N(N-1)/2}\left[\pi_{\mathcal L}^{(m,s)}(\Lambda)^\dagger\right]^{\otimes N}.
\end{equation}
Considering the action of $\pi_{\mathcal L}^{q}(\Lambda)$ on the pair-wise helicity part of the state one gets:
\begin{multline}\label{eq:twirling_pair_wise}
    \int_{\mathcal L}\operatorname{d}\!\Lambda f(\Lambda)\sum_{k,l}c_{n^{(k)}}c_{n^{(l)}}\left[\pi_{\mathcal L}^{(m,s)}(\Lambda)\right]^{\otimes N}\ket{p^N; \sigma^N_{n^{(k)}}}\bra{p^N; \sigma^N_{n^{(l)}}}\left[\pi_{\mathcal L}^{(m,s)}(\Lambda)^\dagger\right]^{\otimes N}\\
    \otimes\ket{(p^N),\lbrace q_{{\alpha}{\alpha'}}^k\rbrace}\bra{(p^N),\lbrace q_{{\alpha}{\alpha'}}^l\rbrace}e^{i\left(\sum_{{\alpha}>{\alpha'}}(q_{{\alpha}{\alpha'}}^{k}-q_{{\alpha}{\alpha'}}^{l})\right)\phi_{\Lambda}}.
\end{multline}
\end{widetext}
Here we have used the fact that due to the assumption on equal momenta for particles of a given species for $q_{\alpha\alpha'}\neq0$   all $\phi_{\Lambda,{\alpha}{\alpha'}}(p_\alpha,p_{\alpha'})$ are equal to some $\phi_{\Lambda}$ as in this case the pair of momenta parameterizing the phase is always the same as it corresponds to two particles of different species. Additionally, whenever $q_{\alpha\alpha'}=0$ value of $\phi_{\Lambda,{\alpha}{\alpha'}}(p_\alpha,p_{\alpha'})$ is irrelevant and can be put to be equal to $\phi_{\Lambda}$. In analogy to the scenario for massless particles, if one restricts states to fulfill:
\begin{equation}\label{eq:sum_q}
\forall_k\,\, \sum_{{\alpha}>{\alpha'}}q^{k}_{{\alpha}{\alpha'}}=q,
\end{equation}
then the phase factor disappears. Additionally, if the state $\rho^{p,q,m_{e,g}>0}$ was chosen to be separable in the virtual partition between the spin-momentum part and pair-wise helicity then one can see that the reduced state of the pair-wise helicity becomes unaltered by the twirling operation. This is because of the fact that transformation factorizes in this partition and thus leaves separable states separable. To see this directly, let us write the state $\rho^{p,q,m_{e,g}>0}$ as:

\begin{equation}
    \rho^{p,q,m_{e,g}>0}=\sum_t p_t\rho_t^{p,m_{e,g}>0}\otimes\rho_t^{q},
\end{equation}
where $\rho_t^{p,m_{e,g}>0}$ and $\rho_t^{q}$ stand for density matrices in spin-momentum and pair-wise helicity subsystems respectively. Then upon \eqref{eq:twirling_pair_wise} and assuming \eqref{eq:sum_q} we get
\begin{widetext}
    \begin{multline}
    \mathcal T_{\mathcal L}(\rho^{p,q,m_{e,g}>0})=\\\sum_tp_t\int_{\mathcal L}\operatorname{d}\!\Lambda f(\Lambda)\left[\pi_{\mathcal L}^{(m,s)}(\Lambda)\right]^{\otimes N}\rho_t^{p,m_{e,g}>0}\left[\pi_{\mathcal L}^{(m,s)}(\Lambda)^\dagger\right]^{\otimes N}\otimes\left[\pi_{\mathcal L}^{q}(\Lambda)\right]^{\otimes N(N-1)/2}\rho_t^{q}\left[\pi_{\mathcal L}^{q}(\Lambda)^{\dagger}\right]^{\otimes N(N-1)/2}\\
    =\sum_tp_t\mathcal T_{\mathcal L}(\rho_t^{p,m_{e,g}>0})\otimes\rho_t^{q}.
\end{multline}
\end{widetext}
Clearly, the resulting state is also separable with the unaltered pair-wise helicity part. Thus, in such a scenario, any information encoded solely in the pair-wise helicity is not impacted by Poincar\'e transformations. One can still choose the spin part of the state to be relativistically invariant, as discussed in the previous sections, making all information encoded in the state  relativistically invariant. It is however important to notice that due to the different masses, Lorentz transformations do not act collectively on all spin degrees of freedom but the action is collective only on the parts of the state corresponding to given species of particles.  Therefore, the equal momentum encoding scheme described in previous sections should be realised separately within the sets of spins corresponding to different species of particles. Another option is to utilise the non-equal momentum encoding described in the next section.

Finally, let us analyze how one can obtain a relativistically invariant encoding upon the simple encoding presented in the beginning of the section. Consider two encoding pairs (see Fig. \ref{fig:1}), one with pair-wise helicity equal to $1$ and the other one equal to $-1$. Those pairs comprise  the encoding cell to which one can assign bit 1 when $q_{12}=1, q_{34}=-1$ and bit $0$ when  $q_{12}=-1, q_{34}=1$. It is easy to find that $\sum_{{\alpha}>{\alpha'}}q_{{\alpha}{\alpha'}}=0$ for both situations. Therefore, this encoding is relativistically invariant. Now, one has to consider if inclusion in the state of another cell will not destroy this property as the pair-wise helicities will appear between all particles. To this matter, it is enough to check sum of pair-wise helicities between pairs with configurations of pair-wise helicities ($1,1$) and ($-1,-1$) as configurations (1,-1) and (-1,1) where already considered in the analysis of the encoding cell. As sum of pair-wise helicities between such pairs turns out to be also equal to 0 (see Fig. \ref{fig:1}) including additional cells which encode $1$ and $0$ as described above will not affect the total sum of $q_{{\alpha}{\alpha'}}$ and therefore the relativistic invariance will be achieved.

\subsection{Non-equal momentum encodings} \label{sectionIII_non_col}
\label{sec:NonColMoms}
In all relativistically invariant encodings of quantum information proposed so far in the literature \cite{Bartlett05} one assumes that all the particles used for encoding have the same almost sharp value of momentum, as in this case the Wigner rotation acts identically on all the particles involved. However, the assumption of \textit{equal momenta of all the particles} can be traded for an assumption of \textit{fixed total momentum} by utilizing multiparticle irreducible representations \eqref{PoincareIrreps} of the Poincare group. The states $\ket{\mathfrak{s},J;P,\Sigma_J,\{\lambda^{\textrm{CM}}_{\alpha}\}}$ for fixed values of total spin $J$ and centre-of-mass-frame energy $\mathfrak{s}$ span irreducible subspaces under the action of the Poincare group.  They are labelled by the total momentum $P$, total helicity $\Sigma_J$ and the set of helicities $\{\lambda^{\textrm{CM}}_{\alpha}\}$ of all the particles in the centre of mass frame, which serve as multiplicity labels. In the following, we will show that these states with fixed total momentum $P$ can be used for an encoding of quantum information invariant with respect to arbitrary Lorentz transformations. We will proceed in two steps, (i) firstly we will analyze transformation properties of such states focusing on helicity states, (ii) secondly we will show how to construct an invariant encoding using the multiplicity subspaces. 
\begin{enumerate}[label=(\roman*)]
    \item For simplicity of notation let us denote the $k$-th element of the set of helicities by $\mu_k$, that is $\mu_k=\{\lambda^{\textrm{CM}}_{\alpha}\}_k$. Then let us consider an arbitrary state of the form:
    \begin{equation}
        \label{TotHelState}
        \ket{\Psi}^{\mathfrak{s},J,P}=\sum_{\Sigma_J=-J}^J\sum_{k} c_{\Sigma_J,\mu_k}\ket{\mathfrak{s},J;P,\Sigma_J,\mu_k},
    \end{equation}
    and its corresponding density matrix by $\rho^{\mathfrak{s},J,P}$. Then the twirling operation with respect to the Lorentz group applied to this state reads:
    \begin{eqnarray}
         \label{PoincareTwirlMass}
    &&\mathcal T_{\mathcal L}(\rho^{\mathfrak{s},J,P})=\nonumber\\
    &&\int_{\mathcal L}\operatorname{d}\!\Lambda f(\Lambda)\pi_{\mathcal L}^{(\mathfrak{s},J)}(\Lambda)\rho^{\mathfrak{s},J,P} \pi_{\mathcal L}^{(\mathfrak{s},J)}(\Lambda)^{\dagger},\nonumber\\
    \end{eqnarray}
in which $\pi_{\mathcal L}^{(\mathfrak{s},J)}(\Lambda)=\pi_{\mathcal P}^{(\mathfrak{s},J)}(\Lambda,0)$ is an irreducible representation of Poincare group \eqref{PoincareIrreps} restricted to the Lorentz group, reads as follows:
\begin{eqnarray}
\label{spinTwirlTotalMom}
     &&\mathcal T_{\mathcal L}(\rho^{\mathfrak{s},J,P})=\int_{\mathcal L}\operatorname{d}\!\Lambda f(\Lambda)N_{\Lambda,p}^2\sum_{\Sigma_J,k}\sum_{\Sigma_J',k'}c_{\Sigma_J,\mu_k}c_{\Sigma_J',\mu_k'}^*\nonumber\\
     &&\operatorname{D}^{(J)}(\Omega_{\Lambda}^P)\ket{\mathfrak{s},J;\Lambda P,\Sigma_J,\mu_k}\bra{\mu_k',\Sigma_J',\Lambda P;J,\mathfrak{s}}\operatorname{D}^{(J)}(\Omega_{\Lambda}^P)^{\dagger}.\nonumber\\
\end{eqnarray}
Due to the fact that the Wigner rotation acts trivially on the multiplicity degrees of freedom, one can utilize them for invariant encoding.
\item In the second step we show how to perform an invariant encoding in situation described by \eqref{spinTwirlTotalMom}. 
 We assume that the values of the total momentum $P$ and total rest energy $\mathfrak{s}$ are fixed. Then Lorentz transformations $\Lambda$ act via Wigner rotation in the same way on each of the  states $\ket{\mathfrak{s},J;P,\Sigma_J,\mu_k}$, hence we can utilize the encoding scheme described in Section \ref{sec:genInvState} with the following identification of the Schur basis elements: $i\mapsto J$, $r \mapsto \Sigma_J$, $\mu\mapsto\mu_k$.  Then the states $\ket{\mathfrak{s},J;P,\Sigma_J,\mu_k}$ can be for each value of $J$ organized into arrays analogous to \eqref{SchurBasis}.  The upper bounds on dimensions $D_L^J$ and $D_V^J$ can be easily evaluated to $D^J_L\leq 2J+1$ and $D^J_V\leq |\lambda^{\textrm{CM}}|^N$, where $|\lambda^{\textrm{CM}}|$ denotes the number of different possible values of helicity for each of the particles whereas $N$ denotes total number of particles . However in specific situations both dimensions can be strictly lower due to vanishing of respective Poincare-Clebsch-Gordan coefficients in \eqref{partialWaveInt}.
In order to introduce an invariant encoding it is advisable to separate momentum and angular momentum parts of the state in full analogy with the case of equal momenta encoding \eqref{spinTwirl2}:
\begin{eqnarray}
     &&\mathcal T_{\mathcal L}(\rho^{\mathfrak{s},J,P})=\int_{\mathcal L}\operatorname{d}\!\Lambda f(\Lambda)N_{\Lambda,p}^2\ket{\mathfrak{s}; \Lambda P}\bra{\Lambda P; \mathfrak{s}}\otimes\sum_{\Sigma_J,k}\sum_{\Sigma_J',k'}\nonumber\\
     &&c_{\Sigma_J,\mu_k}c_{\Sigma_J',\mu_k'}^*\operatorname{D}^{(J)}(\Omega_{\Lambda}^P)\ket{J;\Sigma_J,\mu_k}\bra{\mu_k',\Sigma_J';J}\operatorname{D}^{(J)}(\Omega_{\Lambda}^P)^{\dagger}.\nonumber\\
\end{eqnarray}
In full analogy with \eqref{inStateMassive} we take an initial state in the form:
\begin{equation}
    \label{inStateIrrep}
    \rho_{\mathcal L}^{\mathfrak{s},J,P}=\left(\ket{\mathfrak{s}; P}\bra{P; \mathfrak{s}}\right)\otimes \rho^{\mathfrak{s},J}_{\mathcal L},
\end{equation}
in which $\rho^{\mathfrak{s},J}_{\mathcal L}$ reads:
\begin{equation}
    \label{rhoInvariantJSigma}
    \rho_{\mathcal L}^{\mathfrak{s},J}=\frac{1}{D^J_L}\sum_{\mu_1\mu_2=1}^{D^J_V}\rho^{\mathfrak{s},J}_{\mu_1\mu_2}\hat\Pi_{J}^{\mu_1\mu_2}, 
\end{equation}
and the operators $\hat\Pi^{\mu_1\mu_2}_J$ are defined according to \eqref{PiBasis} by:
\begin{equation}
    \hat\Pi^{\mu_1\mu_2}_J=\sum_{\Sigma_J=1}^{D^J_L}\ket{J;\Sigma_J,\mu_1}\bra{\mu_2,\Sigma_J;J}.
\end{equation}
The spin part of the state $\rho_{\mathcal L}^{\mathfrak{s},J,P}$ is invariant with respect to the action of the Lorentz group. Indeed, in analogy with \eqref{spinTwirl3}, for arbitrary probability measure $\operatorname{d}\!\Lambda f(\Lambda)$ we have:
\begin{eqnarray}
    \mathcal T_{\mathcal L}(\rho^{\mathfrak{s},J,P})&=&\int_{\mathcal L}\operatorname{d}\!\Lambda f(\Lambda)N_{\Lambda,p}^2\ket{\mathfrak{s}; \Lambda P}\bra{\Lambda P; \mathfrak{s}}\otimes\nonumber\\
    &&\operatorname{D}^{(J)}(\Omega_{\Lambda}^P)\rho^{\mathfrak{s},J}_{\mathcal L}\operatorname{D}^{(J)}(\Omega_{\Lambda}^P)^{\dagger}\nonumber\\
    &=&\int_{\mathcal L}\operatorname{d}\!\Lambda f(\Lambda)N_{\Lambda,p}^2\ket{\mathfrak{s}; \Lambda P}\bra{\Lambda P; \mathfrak{s}}\otimes\rho^{\mathfrak{s},J}_{\mathcal L}\nonumber\\
     &&\operatorname{D}^{(J)}(\Omega_{\Lambda}^P)\operatorname{D}^{(J)}(\Omega_{\Lambda}^P)^{\dagger}\nonumber\\
     &=&\left(\int_{\mathcal L}\operatorname{d}\!\Lambda f(\Lambda)N_{\Lambda,p}^2\ket{\mathfrak{s}; \Lambda P}\bra{\Lambda P; \mathfrak{s}}\right)\otimes\rho^{\mathfrak{s},J}_{\mathcal L}.\nonumber\\
\end{eqnarray}
Note that in our encoding the values of the total centre-of-mass-frame energy $\mathfrak{s}$, total angular momentum $J$ and of the total momentum $P$ are fixed.

\end{enumerate}

Let us conclude this section with two remarks. Firstly, the invariant encoding scheme based on equal momenta for massive particles is a specific example of the above presented fixed total momentum encoding, in which all the total angular momentum $J$ comes from adding spin angular momentum. Indeed, if all the particles have fixed and equal momentum, their total momentum is also fixed as well as all relative momenta in the centre of mass frame. Therefore, integration in \eqref{partialWaveInt} trivializes and there is no additional orbital angular momentum component that contributes to the total angular momentum $J$. In the case of massless particles, the case of equal momentum encoding is not included in the fixed total momentum encoding scheme, since there is no well-defined centre of mass frame for such a system.

Secondly, note that in the case of fixed total momentum encoding one must assume that encoding is performed using states with sharp total momentum, which means that the centre of mass position is totally delocalized. Therefore, in reality, one can fulfill this condition only to a reasonable extent, in analogy to the case of equal momentum encoding assuming sharp momenta of all the particles.

\section{Conclusions} \label{conclusions}

The current theory of relativistically invariant encodings still has its shortcomings and is not fully general.  Here we address these problems, and to this matter we have presented the full theory of invariant encoding under Lorentz transformations. We have extended previous results by proposing non-equal momentum encoding which does not require equal momentum of all particles encoding information. This approach retrieves earlier results when one assumes equal momentum of all particles. We also considered the full classification of representations of  Poincar\'e group which includes the appearance of transformations of pair-wise helicity in multiparticle states which appear non-trivially in systems with electric and magnetic charges. This allowed us to propose encoding using pair-wise helicity which encodes more information per particle, than previously considered schemes. What is more, we provide a derivation which is rigorus and transparent for existence of invariant encodings for both compact and non-compact transformation groups acting via unitary representations, which is valid for any distribution of group transformations and not only for uniform distributions. This is important since it is a custom to consider scenario of complete ''lack of knowledge`` about reference frame, which is justified e. g. for phase reference of optical beam as each pulse of the laser can result in different random phase; however, it is not justified in the case of Lorentz boosts effectively decohering states. Therefore, it is important to state clearly that Decoherence-Free Subspaces and Decoherence-Free Subsystems work for any distribution  of the decohering group operations.

However, the theory of relativistically invariant encoding, while highly developed, is still far from being completed. In our considerations, we have only used special relativity, while curved spacetimes can, in fact, significantly impact encoded quantum information \cite{CSP}. This is relevant, as relativistically invariant encoding is also intended for use in space communication where the changes of gravitational field are unavoidable. Therefore, one could ask how the curved spacetime affects the non-equal momentum encoding proposed by us.
\section{Acknowledgements}

We acknowledge support by the Foundation for Polish Science (IRAP project, ICTQT, contract no. MAB/2018/5, co-financed by EU within Smart Growth Operational Programme).

\bibliography{helicity}

\appendix 
\section{Invariant encoding in the case of non-compact symmetry group acting on the state space via non-unitary representations.}
\label{app:SLOCC}
Let us take as an example the case of $\mathcal G=\textrm{SL}(d, \mathbb C)$, and the following representation of $\textrm{SL}(d, \mathbb C)$ on $(\mathbb C^d)^{\otimes N}$ based on its Cartan decomposition. Let us first recall that arbitrary element $M$ of $\textrm{SL}(d, \mathbb C)$ can be (non-uniquely) represented as $M=KAK'$, where $K,K'$ are special unitary $\textrm{SU}(d)$ matrices, whereas $A$ is some diagonal real matrix of determinant one. Then consider the following representation:
\begin{equation}
    \label{sl-rep}
    \pi_{\operatorname{SL}(d, \mathbb C)}(M)=\left(KA_{\operatorname{n}}K'\right)^{\otimes N},
\end{equation}
in which $A_{\operatorname{n}}=A/||A||$ and the corresponding product measure $\operatorname{d}M=\operatorname{d}\mu(K)\operatorname{d}\mu(A)\operatorname{d}\mu(K')$, in which $\operatorname{d}\mu(K)$, $\operatorname{d}\mu(K')$ are normalised Haar measures on $\textrm{SU}(d)$, whereas $\operatorname{d}\mu(A)$ is some normalised measure on a non-compact manifold corresponding to the maximal Abelian subgroup of  $\textrm{SL}(d, \mathbb C)$. Then it can be shown that the map \eqref{def-twirl-gen} represents averaging of $d$-dimensional quantum states over collective noise represented by SLOCC (Stochastic Local Operations and Classical Communication) operations. It turns out that the commutant of $\pi_{\operatorname{SL}(d, \mathbb C)}(\operatorname{SL}(d, \mathbb C))$ is once more the permutation group $\operatorname{S}_N$, and that the action of \eqref{def-twirl-gen} for the case of $\mathcal G=\operatorname{SL}(d, \mathbb C)$ on the states invariant with respect to the collective action of the unitary group just rescales the states corresponding to the \textit{invariant subspaces} with respect to the collective action of the unitary group. Therefore, if we perform postselection onto one of such invariant subspaces, the quantum information encoded in this subspace is also invariant with respect to the collective action of $\operatorname{SL}(d, \mathbb C)$ -- collective SLOCC operations. 

\end{document}